\documentclass[a4paper,11pt]{article}
\pdfoutput=1
\usepackage{jheppub}
\usepackage[numbers,sort&compress]{natbib}
\usepackage{graphicx}  % needed for figures
\usepackage{float}
\usepackage{dcolumn}   % needed for some tables
\usepackage{bm}        % for math
\usepackage{amssymb}   % for math
\usepackage{slashed}   % for Dirac Slash
\usepackage{amsmath}   % for mutiline eqn
\usepackage{verbatim}  % for multi-line comment
\usepackage{color} % for textcolor
\usepackage{epstopdf}

\definecolor{Blue}{rgb}{0, 0.1, 0.5}

\newcounter{YJC}

\usepackage{hyperref}
\graphicspath{{fig/}}
\begin{document}
%\widetext

\title{Measuring the anomalous quartic gauge couplings in the $W^+W^-\to W^+W^-$ process at muon collider using artificial neural networks}

\author[a]{Ji-Chong Yang}
\emailAdd{yangjichong@lnnu.edu.cn}
\affiliation[a]{Department of Physics, Liaoning Normal University, Dalian 116029, China}

\author[a]{Xue-Ying Han}

\author[a]{Zhi-Bin Qin}

\author[b]{Tong Li}
\emailAdd{litong@nankai.edu.cn}
\affiliation[b]{School of Physics, Nankai University, Tianjin 300071, China}

\author[a]{Yu-Chen Guo}
\emailAdd{ycguo@lnnu.edu.cn}
\thanks{Corresponding author}

\abstract{
The muon collider provides a unique opportunity to study the vector boson scattering processes and dimension-8 operators contributing to anomalous quartic gauge couplings~(aQGCs).
Because of the cleaner final state, it is easier to decode subprocess and certain operator couplings at a muon collider.
We attempt to identify the anomalous $WWWW$ coupling in the exclusive $WW\to WW$ scattering in this paper.
Since one aQGC can be induced by multiple dimension-8 operators, the study of one coupling can help to confine different operators.
Meanwhile, singling out the $WW\to WW$ process can help to study the unitarity bounds.
The vector boson scattering process corresponding to the anomalous $WWWW$ coupling is $\mu^+\mu^-\to \nu\nu\bar{\nu}\bar{\nu}\ell^+\ell^-$, with four (anti-)neutrinos in the final state, which brings troubles in phenomenological studies.
In this paper, the machine learning method is used to tackle this problem.
We find that, using the artificial neural network can extract the $W^+W^-\to W^+W^-$ contribution, and is helpful to reconstruct the center of mass energy of the subprocess which is important in the study of the Standard Model effective field theory.
The sensitivities and the expected constraints on the dimension-8 operators at the muon collider with $\sqrt{s}=30$ TeV are presented.
We demonstrate that the artificial neural networks exhibit great potential in the phenomenological study of processes with multiple neutrinos in the final state.
%\end{abstract}
}

\maketitle

\section{\label{sec1}Introduction}

%The vector boson scattering~(VBS) processes are currently one kind of the most talked about processes at the Large Hadron Collider~(LHC).
The self-couplings of electroweak (EW) gauge bosons are most closely related to the nature of the electroweak symmetry breaking (EWSB)~\cite{vbs1,vbs2,vbscan,positivity1,positivity2}. Any hints for the anomalous gauge couplings would indicate the existence of new physics (NP) beyond the Standard Model (SM). In the framework of the SM effective field theory~(SMEFT)~\cite{weinberg,SMEFTReview1,SMEFTReview2,SMEFTReview3}, the dimension-8 operators contribute to the anomalous quartic gauge couplings~(aQGCs)~\cite{aqgcold,aqgcnew}.
On the other hand, the vector boson scattering~(VBS) is one of the most common channels for performing precision measurement of the SM or searching NP beyond the SM at high-energy colliders.
The probe of aQGCs through VBS is thus one of the most important topics at the Large Hadron Collider (LHC) and has received great attention~\cite{sswwexp1,sswwexp2,zaexp1,zaexp2,zaexp3,waexp1,zzexp1,zzexp2,wzexp1,wzexp2,wwexp1,wwexp2,wvzvexp,waexp2,zzexp3}. Nevertheless, the VBS measurements suffer from the large QCD background at the LHC and it is difficult to decode the initial states of subprocess as the final jets in the forward region are not distinguishable.

%Due to the delicate cancellation from the Higgs boson which is assumed to be extremely susceptible to new physics~(NP) beyond the Standard Model~(SM), the VBS processes are considered to be excellent windows to probe NP affecting electric-weak symmetry broken~(EWSB)~\cite{vbs1,*vbs2,*vbscan,positivity1,*positivity2}.
%EWSB-related measurements still provide a rigorous test of the SM until today, such as the recently discovered $W$ boson mass problem~\cite{wmass}.
%In the framework of the SM effective field theory~(SMEFT)~\cite{weinberg,*SMEFTReview1,*SMEFTReview2,*SMEFTReview3}, the dimension-8 operators contributing to the anomalous quartic gauge couplings~(aQGCs)~\cite{aqgcold,*aqgcnew} are just as suitable to be investigated with VBS processes.
%Therefore, and the VBS as well as aQGCs have received great attention~\cite{sswwexp1,*sswwexp2,*zaexp1,*zaexp2,*zaexp3,*waexp1,*zzexp1,*zzexp2,*wzexp1,*wzexp2,*wwexp1,*wwexp2,wvzvexp,waexp2,zzexp3}.

Recently, the development of the muon collider has gradually entered the limelight~\cite{muoncollider1,muoncollider2,muoncollider4,muoncollider6,muoncollider7,muoncollider8,muoncollider3,muoncollider5}.
On the high-energy muon collider, the dominant production mode for the SM and NP particles is VBS or vector boson fusion process~\cite{muoncollider3}.
Therefore, the muon collider is also known as a gauge boson collider~\cite{muoncollider5}.
Compared to the LHC, there are no composite particles in the initial states at muon collider, and thus the QCD background is not severe.
Taking $WW$ initiated scattering as an example, the high-energy muon beams radiate $W$ bosons and turn into neutrinos. Meanwhile, the neutral gauge bosons $Z,\gamma$ are also radiated under an approximately unbroken SM gauge symmetry and muons are produced in final states. The outgoing muons are extremely forward with a small polar angle of the order $\theta_\mu\sim M_Z/E_\mu\approx 1.2^\circ$ for a $Z$-initiated process at 10 TeV~\cite{Han:2020pif} and most likely escape the detector. If we require the outgoing muons to be observable in the detector coverage $10^\circ < \theta_\mu < 170^\circ$, the cross sections of neutral gauge bosons initiated scattering would be substantially suppressed by two orders of magnitude~\cite{Han:2020uak,Han:2022mzp}.
Thus, it is feasible to study the aQGCs induced exclusive $W^+W^- \to W^+W^-$ scattering at a muon collider.
In addition, the muon collider can reach both high energy and high luminosity, which will be of great help to precisely measure aQGCs, since the cross-section induced by dimension-8 operators increases significantly with energy.
Meanwhile, high luminosity is considered as one of the keys to solve the ``EFT triangle'' problem~\cite{mo1,efttraingle2,efttraingle3,wwwwunitary}.

%Taking aQGCs as an example, it is difficult to separate $W^+W^-\to W^+W^-$ from $VV\to W^+W^-$ at the LHC.
%However, on a muon collider, this will become possible, for example, as will be shown the $W^+W^-\to W^+W^-$ process contributed by the scalar/longitudinal operators is dominant in the contribution of NP in process $\mu^+\mu^-\to \nu\nu\bar{\nu}\bar{\nu}\ell^+\ell^-$.
%In addition to this, the muon collider can touch the boundary of both high energy and high luminosity, which will be of great help to study aQGCs, since the cross-section of the dimension-8 operators increases significantly with energy.
%Meanwhile, high luminosity is considered as one of the keys to solve the `EFT triangle' problem~\cite{mo1,*efttraingle2,*efttraingle3,wwwwunitary}.

In this work, we investigate the sensitivity of $W^+W^-\to W^+W^-$ scattering to the dimension-8 scalar/longitudinal operators contributing to aQGCs at muon colliders.~\footnote{A recent Snowmass paper investigated the searches of aQGCs through the production of $WW$ boson pairs at a muon collider with $\sqrt{s}=6$ TeV and an integrated luminosity of 4 ab$^{-1}$~\cite{Abbott:2022jqq}. They studied the $WW\nu\nu$ and $WW\mu\mu$ final states with the $W$ bosons decaying hadronically.} The $W$ bosons in final states are then followed by purely leptonic decay $W^\pm\to \ell^\pm \nu$.
One key problem of the process $\mu^+\mu^-\to \nu\nu\bar{\nu}\bar{\nu}\ell^+\ell^-$ is the presence of the (anti-)neutrinos which lead to difficulties in the phenomenological studies.
For example, it is difficult to reconstruct the center of mass~(c.m.) energy of the subprocess $W^+W^-\to W^+W^-$~(denoted as $\sqrt{\hat{s}}$).
In the content of EFT, the Wilson coefficients of effective operators should rely on energy scales~\cite{matchingidea1}.
At a high-energy muon collider, the c.m. energy is especially important because at high energies, the validity of SMEFT should be taken into account since the SMEFT is valid only under a certain energy scale.
The unitarity bound~\cite{unitarityHistory1,unitarityHistory2,unitarityHistory3,partialwaveunitaritybound,jrr1} is often needed to investigate the validity of SMEFT, and the c.m. energy is necessary information to apply unitarity bounds.
To solve the problem of reconstructing $\hat{s}$ in the processes with multiple (anti-)neutrinos, a machine learning approach has been introduced into high energy physics (HEP) community~\cite{wwwwunitary}.
The machine learning methods have been widely used, and are being rapidly developed in HEP~\cite{wpolarizationANN1,wpolarizationANN2,zpolarizationANN,taupolarizationANN,annhep1,annhep2,annhep3,mlreview,ml1,ml2,ml3,ml4,ml5,ml6,ml7}.
In this paper, we adopt the artificial neural network~(ANN) to extract the $W^+W^-\to W^+W^-$ contribution and to reconstruct $\hat{s}$.
The complexity caused by the neutrinos just provides a venue to explore the boundaries of ANN capabilities.
Based on the ANNs, the sensitivities of the process $\mu^+\mu^-\to \nu\nu\bar{\nu}\bar{\nu}\ell^+\ell^-$ to the dimension-8 operators contributing to aQGCs are investigated, with the focus on the $W^+W^-\to W^+W^-$ contribution.

The rest of this paper is organized as follows.
In Sec.~\ref{sec2}, the dimension-8 operators contributing to aQGCs are briefly reviewed.
The ANN approach to extract the $W^+W^-\to W^+W^-$ contribution is discussed in Sec.~\ref{sec3}.
In Sec.~\ref{sec4}, we discuss the ANN approach to reconstruct $\hat{s}$.
The expected constraints on the coefficients of the aQGC operators at the muon collider are estimated in Sec.~\ref{sec5}.
Sec.~\ref{sec6} summarizes our main conclusions.

%%%%%%%%%%%%%%%%%%%%%%%%%%%%%%%%%%%%%%
\section{\label{sec2}A brief introduction of the anomalous quartic gauge couplings}
%%%%%%%%%%%%%%%%%%%%%%%%%%%%%%%%%%%%%%

The Lagrangian of the dimension-8 operators contributing to aQGCs can be written as~\cite{aqgcold,aqgcnew}
\begin{equation}
\begin{split}
&\mathcal{L}_{\rm aQGC}=\sum _{i=0}^2 \frac{f_{S_i}}{\Lambda^4}O_{S,i}+\sum _{j=0}^7 \frac{f_{M_j}}{\Lambda^4}O_{M,j}+\sum _{k=0}^9 \frac{f_{T_k}}{\Lambda^4}O_{T,k}
\end{split}
\label{eq.2.1}
\end{equation}
with
\begin{equation}
\begin{split}
&O_{S,0}=\left[\left(D_{\mu}\Phi \right) ^{\dagger} D_{\nu}\Phi\right]\times \left[\left(D^{\mu}\Phi \right) ^{\dagger} D^{\nu}\Phi\right],\\
&O_{S,2}=\left[\left(D_{\mu}\Phi \right) ^{\dagger} D_{\nu}\Phi\right]\times \left[\left(D^{\nu}\Phi \right) ^{\dagger} D^{\mu}\Phi\right],\\
\end{split}
\quad
\begin{split}
&O_{S,1}=\left[\left(D_{\mu}\Phi \right) ^{\dagger} D_{\mu}\Phi\right]\times \left[\left(D^{\nu}\Phi \right) ^{\dagger} D^{\nu}\Phi\right],\\
&\\
\end{split}
\label{eq.2.2}
\end{equation}
\begin{equation}
\begin{split}
&O_{M,0}={\rm Tr\left[\widehat{W}_{\mu\nu}\widehat{W}^{\mu\nu}\right]}\times \left[\left(D_{\beta}\Phi \right) ^{\dagger} D^{\beta}\Phi\right],\\
&O_{M,2}=\left[B_{\mu\nu}B^{\mu\nu}\right]\times \left[\left(D_{\beta}\Phi \right) ^{\dagger} D^{\beta}\Phi\right],\\
&O_{M,4}=\left[\left(D_{\mu}\Phi \right)^{\dagger}\widehat{W}_{\beta\nu} D^{\mu}\Phi\right]\times B^{\beta\nu},\\
&O_{M,7}=\left(D_{\mu}\Phi \right)^{\dagger}\widehat{W}_{\beta\nu}\widehat{W}^{\beta\mu} D^{\nu}\Phi,\\
\end{split}
\quad
\begin{split}
&O_{M,1}={\rm Tr\left[\widehat{W}_{\mu\nu}\widehat{W}^{\nu\beta}\right]}\times \left[\left(D_{\beta}\Phi \right) ^{\dagger} D^{\mu}\Phi\right],\\
&O_{M,3}=\left[B_{\mu\nu}B^{\nu\beta}\right]\times \left[\left(D_{\beta}\Phi \right) ^{\dagger} D^{\mu}\Phi\right],\\
&O_{M,5}=\left[\left(D_{\mu}\Phi \right)^{\dagger}\widehat{W}_{\beta\nu} D^{\nu}\Phi\right]\times B^{\beta\mu} + h.c.,\\
&\\
\end{split}
\label{eq.2.3}
\end{equation}
\begin{equation}
\begin{split}
&O_{T,0}={\rm Tr}\left[\widehat{W}_{\mu\nu}\widehat{W}^{\mu\nu}\right]\times {\rm Tr}\left[\widehat{W}_{\alpha\beta}\widehat{W}^{\alpha\beta}\right],\\
&O_{T,2}={\rm Tr}\left[\widehat{W}_{\alpha\mu}\widehat{W}^{\mu\beta}\right]\times {\rm Tr}\left[\widehat{W}_{\beta\nu}\widehat{W}^{\nu\alpha}\right],\\
&O_{T,6}={\rm Tr}\left[\widehat{W}_{\alpha\nu}\widehat{W}^{\mu\beta}\right]\times B_{\mu\beta}B^{\alpha\nu},\\
&O_{T,8}=B_{\mu\nu}B^{\mu\nu}\times B_{\alpha\beta}B^{\alpha\beta},\\
\end{split}
\quad
\begin{split}
&O_{T,1}={\rm Tr}\left[\widehat{W}_{\alpha\nu}\widehat{W}^{\mu\beta}\right]\times {\rm Tr}\left[\widehat{W}_{\mu\beta}\widehat{W}^{\alpha\nu}\right],\\
&O_{T,5}={\rm Tr}\left[\widehat{W}_{\mu\nu}\widehat{W}^{\mu\nu}\right]\times B_{\alpha\beta}B^{\alpha\beta},\\
&O_{T,7}={\rm Tr}\left[\widehat{W}_{\alpha\mu}\widehat{W}^{\mu\beta}\right]\times B_{\beta\nu}B^{\nu\alpha},\\
&O_{T,9}=B_{\alpha\mu}B^{\mu\beta}\times B_{\beta\nu}B^{\nu\alpha},\\
\end{split}
\label{eq.2.4}
\end{equation}
where $\Phi$ denotes the SM Higgs doublet, $D_{\mu}$ is covariant derivative, $\widehat{W}\equiv \vec{\sigma}\cdot {\vec W}/2$ with $\sigma$ being the Pauli matrix and ${\vec W}=\{W^1,W^2,W^3\}$, $B_{\mu}$ and $W_{\mu}^i$ are $U(1)_Y$ and $SU(2)_I$ gauge fields, and $B_{\mu\nu}$ and $W_{\mu\nu}$ correspond to the gauge invariant field strength tensor.
Many NP models can generate effective dimension-8 operators contributing to aQGCs~\cite{composite1,composite2,extradim,2hdm1,2hdm2,zprime1,zprime2,alp1,alp2,wprime}.
Although the dimension-6 operators have received most studies, the importance of dimension-8 operators was recently emphasized by many groups~\cite{d81,vbs1,ssww,wastudy,zastudy,wwstudy,zastudy,bi1,bi2,bi3,ntgc1,ntgc2,ntgc3,ntgc4,ntgc5,ntgc6,ntgc7}.
As VBS processes receive great attention at the LHC, the above operators in the SMEFT have been investigated intensively.
The LHC constraints on the coefficients of the operators assuming one operator at a time are listed in Table~\ref{tab.1}.
Note that a UV completion model usually does not contribute to only one operator.
The assumption that only one operator exists at a time can be used to study the sensitivity of a process and place stringent constraints when NP beyond the SM has not yet been found.
The scattering process $W^+W^-\to W^+W^-$ can be contributed by $O_{S_{0,1,2}}$, $O_{M_{0,1,7}}$ and $O_{T_{0,1,2}}$ operators, therefore, in the following we concentrate on these operators.

\begin{table}
\begin{center}
\begin{tabular}{c|c||c|c}
\hline
coefficient & constraint & coefficient & constraint \\
\hline
$f_{S_0}/\Lambda^4$ & $[-2.7, 2.7]$~\cite{wvzvexp} & $f_{T_0}/\Lambda ^4$ & $[-0.12, 0.11]$~\cite{wvzvexp} \\
$f_{S_1}/\Lambda^4$ & $[-3.4, 3.4]$~\cite{wvzvexp} & $f_{T_1}/\Lambda ^4$ & $[-0.12, 0.13]$~\cite{wvzvexp} \\
$f_{S_2}/\Lambda^4$ & -                            & $f_{T_2}/\Lambda ^4$ & $[-0.28, 0.28]$~\cite{wvzvexp} \\
$f_{M_0}/\Lambda ^4$ & $[-0.69, 0.70]$~\cite{wvzvexp} & $f_{T_5}/\Lambda ^4$ & $[-0.5, 0.5]$~\cite{waexp2} \\
$f_{M_1}/\Lambda ^4$ & $[-2.0, 2.1]$~\cite{wvzvexp}   & $f_{T_6}/\Lambda ^4$ & $[-0.4, 0.4]$~\cite{waexp2} \\
$f_{M_2}/\Lambda ^4$ & $[-2.8, 2.8]$~\cite{waexp2} & $f_{T_7}/\Lambda ^4$ & $[-0.9, 0.9]$~\cite{waexp2} \\
$f_{M_3}/\Lambda ^4$ & $[-4.4, 4.4]$~\cite{waexp2} & $f_{T_8}/\Lambda ^4$ & $[-0.43, 0.43]$~\cite{zzexp3} \\
$f_{M_4}/\Lambda ^4$ & $[-5, 5]$~\cite{waexp2}     & $f_{T_9}/\Lambda ^4$ & $[-0.92, 0.92]$~\cite{zzexp3} \\
$f_{M_5}/\Lambda ^4$ & $[-8.3, 8.3]$~\cite{waexp2} & & \\
$f_{M_7}/\Lambda ^4$ & $[-3.4, 3.4]$~\cite{waexp2} & & \\
\hline
\end{tabular}
\end{center}
\caption{\label{tab.1}The LHC constraints on the coefficients~(in unit of ${\rm TeV}^{-4}$) of dimension-8 aQGC operators obtained at $95\%$ CL.}
\end{table}

%%%%%%%%%%%%%%%%%%%%%%%%%%%%%%%%%%%%%%%%
\section{\label{sec3}Identification of the \texorpdfstring{$W^+W^-\to W^+W^-$}{W+W- to W+W-} scattering}
%%%%%%%%%%%%%%%%%%%%%%%%%%%%%%%%%%%%%%

%Starting from this section, we neglect the contribution of interference.
%When $\sqrt{s}=30\;{\rm TeV}$, other energy scales are small.
A naive dimensional analysis of the cross-section ignoring inferred divergences and logarithms yields $\sigma _{\rm SM}\sim 1/s$, $\sigma _{\rm int} \sim s f/\Lambda ^4$ and $\sigma _{\rm NP} \sim s^3\left(f/\Lambda ^4\right)^2$, where $\sigma _{\rm SM}$, $\sigma _{\rm int}$ and $\sigma _{\rm NP}$ denote the contributions from the SM, interference term and NP squared term, respectively.
For $\sqrt{s}=30\;{\rm TeV}$, one has $\sigma _{\rm int}\sim \sigma _{\rm NP}$ when $s^2 f/\Lambda ^4 \sim 1$, that is $f/\Lambda ^4 \sim 10^{-6}\;{\rm TeV}^{-4}$. For $f/\Lambda ^4$ above this value, the interference contribution is smaller than $\sigma _{\rm NP}$.
Besides, as the helicity amplitude grow fast with energy, there is not necessarily a corresponding large helicity amplitude in the SM that interferes with NP.
In summary, whether the interference can be neglected needs to be verified for the range of operator coefficients of interest. A numerical justification of this is postponed to Sec.~\ref{sec5.2}.

\begin{figure*}
\begin{center}
\includegraphics[width=0.9\textwidth]{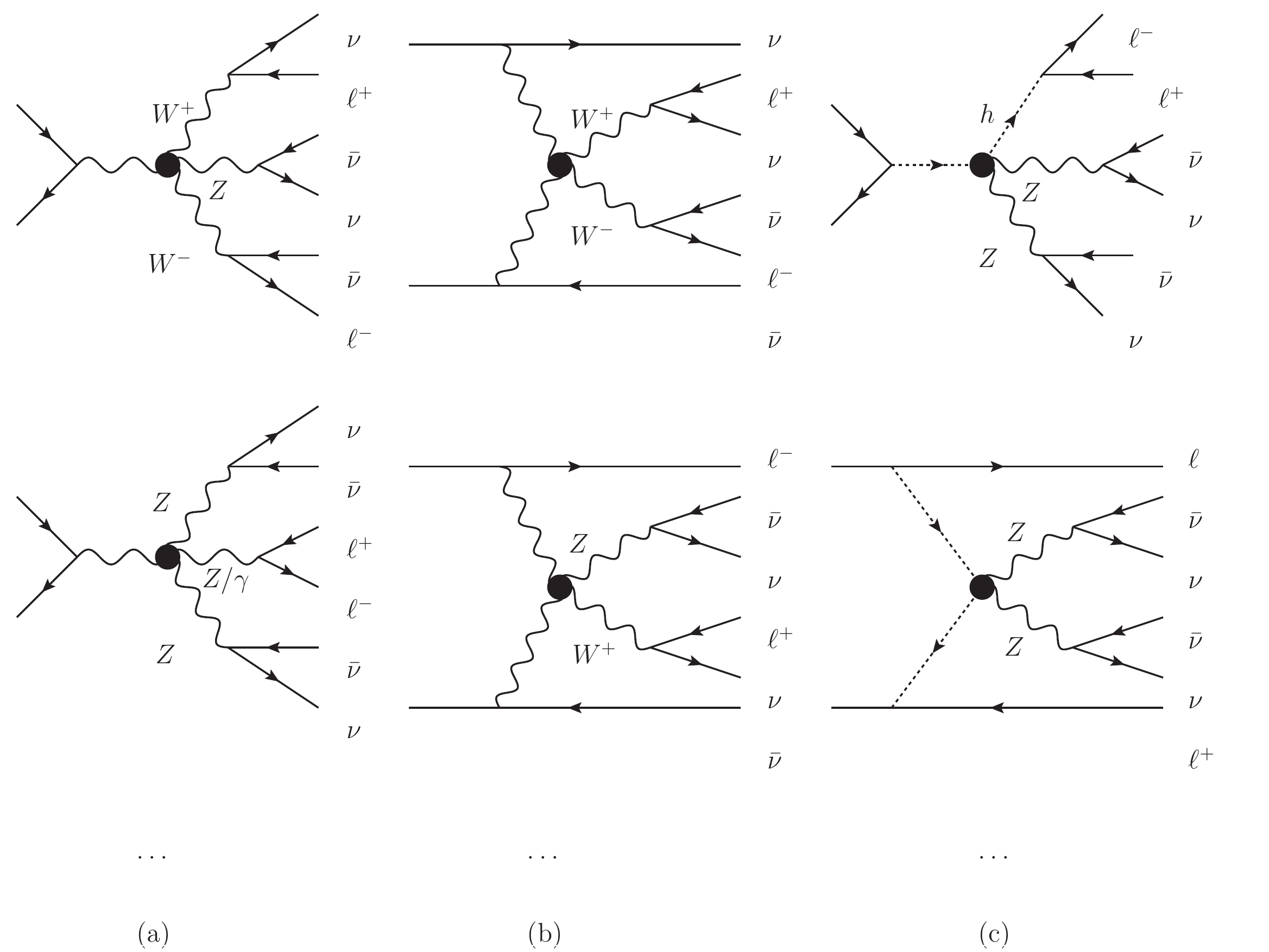}
\caption{\label{fig:diagram}The tree-level Feynman diagrams of the aQGCs contribution to the process $\mu^+\mu^-\to \ell^+\ell^-\nu\nu\bar{\nu}\bar{\nu}$.}
\end{center}
\end{figure*}

The tree-level Feynman diagrams of the aQGCs contribution to the process $\mu^+\mu^-\to \ell^+\ell^-\nu\nu\bar{\nu}\bar{\nu}$ are shown in Fig.~\ref{fig:diagram}.
They can be categorized into three different types, including tri-boson processes~(Fig.~\ref{fig:diagram}~(a)), VBS~(Fig.~\ref{fig:diagram}~(b)) and the Yukawa suppressed diagrams involving Higgs boson~(Fig.~\ref{fig:diagram}~(c)).
Although the VBS processes dominate the contribution of the aQGCs at high energies, there is still significant contribution from the tri-boson process for the above dimension-8 operators~\cite{Yang:2022mie}.
Among the VBS processes, other than $W^+W^-\to W^+W^-$, there are contributions from neutral gauge boson induced processes together with forward muons.
To identify the subprocess $W^+W^-\to W^+W^-$ contribution, it is necessary to select the charged leptons in central region.
In the following, the contribution from diagrams including a $WWWW$ SMEFT interaction is denoted as $\sigma _{\rm 4W}$, and $\sigma _{\rm no-4W}$ denotes other SMEFT contributions.

To study the features of the aQGCs contribution, a Monte Carlo~(M.C.) simulation is applied with the help of the \verb"MadGraph5_aMC@NLO" toolkit~\cite{madgraph,feynrules}.
The events are generated with one operator at a time.
Though out the paper, the standard cuts are set as the default ones in \verb"MadGraph5_aMC@NLO", as
\begin{equation}
\begin{split}
&{\bf p}_{\ell}^T > 10\;{\rm GeV},\;\; |\eta _{\ell}|<2.5,\;\;\Delta R_{\ell\ell}>0.4,\\
\end{split}
\label{eq.2.cuts}
\end{equation}
where ${\bf p}_{\ell}^T$ are the transverse momenta of charged leptons, $\eta _{\ell}$ are the rapidities of charged leptons, and $\Delta R_{\ell\ell}=\sqrt{\Delta \phi^2 + \Delta \eta^2}$ with $\Delta \phi$ and $\Delta \eta$ being the differences of azimuth angles and rapidities of two charged leptons.

Since the neutrinos are invisible, in principle, one cannot use the information of the neutrinos.
However, before we use the machine learning method to identify the contribution of $\sigma _{4W}$, it is useful to illustrate the size of $\sigma _{4W}$ first.
This can provide a criterion for the later algorithms which only utilize detectable observables.
For this purpose, here we temporarily use the information of neutrinos obtained in the M.C. simulations.

Note that there are always two (anti-)neutrinos from a $Z$ boson decay in no-$4W$ processes.
%The flavor of neutrinos can be used to cut off the events which are not from the $W^+W^-\to W^+W^-$ subprocess.
The neutrino flavors from $Z$ boson decay must be the same.
For $W^+W^-\to W^+W^-$, besides the two (anti-)neutrinos $\nu _{\mu}$ and $\bar{\nu}_{\mu}$ along the beam direction, the flavors of the other two neutrinos must correspond to the charged leptons from two $W$ bosons' decay.
We denote $m_{\nu\bar{\nu}}$ as the invariant mass of a pair of (anti-)neutrinos with the same flavor whose invariant mass is closest to $m_Z$ among all possible combinations of neutrinos.
The events are then separated into two groups, according to the neutrino flavors and the size of mass window $\Delta _m=|m_{\nu\bar{\nu}} - m_Z|$.
$\sigma _{\rm 4W}$ is calculated as the cross-section with $\Delta _m > 15\;{\rm GeV}$ and the (anti-)neutrino flavors from $W^+W^-\to W^+W^-$. Meanwhile, $\sigma _{\rm no-4W}$ corresponds to the cross-section with $\Delta _m \leq 15\;{\rm GeV}$ or with wrong flavors.
Their results at $\sqrt{s}=3$, $5$, $10$, $14$ and $30\;{\rm TeV}$~\cite{muoncollider5} are listed in Table~\ref{tab.fraqtion}.

\begin{table}
\begin{center}
\begin{tabular}{c|c|c|c|c|c}
 & $\ell\ell$ & $3$ TeV & $10$ TeV & $14$ TeV & $30$ TeV \\
\hline
$\sigma^{\rm EVA} _{\rm 4W}$ & $ee$ & $0.0048$ & $6.62$ & $49.8$ & $4825.7$ \\
          & $ee$     & $0.00409:0.00016$ & $5.832:0.005$ & $44.14:0.03$ & $4286.7:0.6$ \\
$O_{S_0}$ & $e\mu$   & $0.0078:0.0012$ & $11.7:0.2$ & $88.3:0.8$ & $8571.6:17.4$ \\
          & $\mu\mu$ & $0.0037:0.0015$ & $5.8:0.2$ & $44.0:0.8$ & $4280.5:17.0$ \\
\hline
$\sigma^{\rm EVA} _{\rm 4W}$ & $ee$ & $0.0058$ & $18.9$ & $172.3$ & $24338.5$ \\
          & $ee$     & $0.0061:0.0062$ & $19.2:6.3$ & $173.6:47.1$ & $24044:4515$ \\
$O_{M_0}$ & $e\mu$   & $0.012:0.010$ & $38.4:13.4$ & $346.6:101.2$ & $48302:9812$ \\
          & $\mu\mu$ & $0.0060:0.0092$ & $19.1:10.3$ & $173.6:76.4$ & $24120:7388$ \\
\hline
$\sigma^{\rm EVA} _{\rm 4W}$ & $ee$ & $0.0061$ & $23.7$ & $221.6$ & $32523.9$ \\
          & $ee$     & $0.0069:0.0052$ & $26.8:6.1$ & $250.9:45.6$ & $37083:4618$ \\
$O_{T_0}$ & $e\mu$   & $0.014:0.011$ & $53.7:19.7$ & $502.1:157.8$ & $74450:17396$ \\
          & $\mu\mu$ & $0.0068:0.0117$ & $26.8:19.9$ & $251.6:160.1$ & $37359:17673$ \\
\end{tabular}
\end{center}
\caption{\label{tab.fraqtion}$\sigma _{\rm 4W}:\sigma _{\rm no-4W}$ (fb) for different operators and different charged lepton flavors.
The predictions of effective vector boson approximation~(EVA) are denoted as $\sigma^{\rm EVA} _{\rm 4W}$ for $ee$ final states.
%The predictions of $\sigma _{\rm 4W}$ by effective vector boson approximation~(EVA)~\cite{eva1,eva2,eva3}, which are denoted as $\sigma^{\rm EVA} _{\rm 4W}$~(see Appendix~\ref{ap1}), are also listed as a verification.
%Since EVA is an approximation and there are cuts in M.C. simulations and there is unavoidable interference, a small discrepancy indicates that $\sigma _{\rm 4W}$ is reliable.
}
\end{table}

In Table~\ref{tab.fraqtion}, using the above selection strategy, we show the M.C. results of $\sigma _{\rm 4W}$ and $\sigma _{\rm no-4W}$ given by $O_{S_0}$, $O_{M_0}$ and $O_{T_0}$ for illustration. The Wilson coefficients are taken to be the maximal values allowed by LHC in Table~\ref{tab.1}. The charged leptons in final states are $ee$, $\mu\mu$ or $e\mu$. We also evaluate $\sigma _{\rm 4W}$ with $ee$ charged leptons for comparison, using effective vector boson approximation~(EVA)~\cite{eva1,eva2,eva3}~(denoted as $\sigma _{\rm 4W}^{\rm EVA}$). The detailed EVA calculation is given in Appendix~\ref{ap1}. Although there are kinematic cuts and unavoidable interference in M.C. simulations, the small discrepancy between $\sigma_{\rm 4W}^{\rm EVA}$ and $\sigma_{\rm 4W}$ indicates that our selection strategy of $\sigma _{\rm 4W}$ is reliable.
Table~\ref{tab.fraqtion} shows that if the aQGC signal is induced by $O_{S_i}$ operators, when $\sqrt{s}\geq 10\;{\rm TeV}$, one can concentrate on $\sigma _{\rm 4W}$ and neglect $\sigma _{\rm no-4W}$.
For the cases of $O_{M_i}$ and $O_{T_i}$ operators, although the cross-section of the $W^+W^-\to W^+W^-$ contribution grows with $\sqrt{s}$, $\sigma _{\rm no-4W}$ is not negligible even at $\sqrt{s}= 30\;{\rm TeV}$.
We will use an ANN to select the events from the process involving subprocess $W^+W^-\to W^+W^-$.

Based on the results of Table~\ref{tab.fraqtion}, we find that, to study the VBS subprocess $W^+W^-\to W^+W^-$, a larger $\sqrt{s}$ is needed.
Therefore, we only consider $\sqrt{s}=30\;{\rm TeV}$ below.
In the following, we use $d _{\rm 4W}$ to represent the event data-set from $W^+W^-\to W^+W^-$ contribution, and $d _{\rm no-4W}$ to represent the event data-set from $\sigma _{\rm no-4W}$ contribution.

%%%%%%%%%%%%%%%%%%%%%%%%%%%%%%%%%%%
\subsection{\label{sec3.1}The traditional approach}
%%%%%%%%%%%%%%%%%%%%%%%%%%%%%%%%%%%

\begin{figure}
\begin{center}
\includegraphics[width=0.48\textwidth]{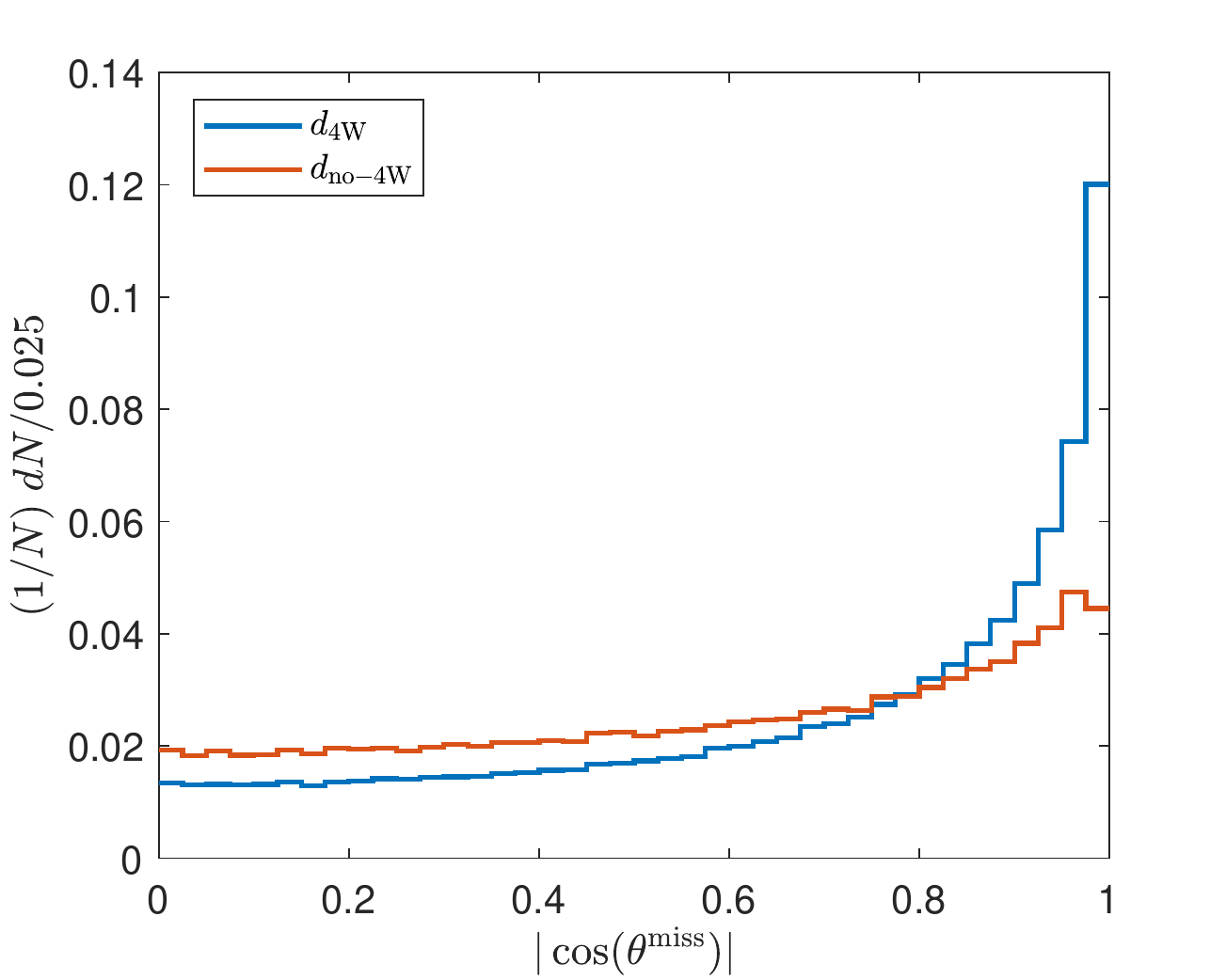}
\includegraphics[width=0.48\textwidth]{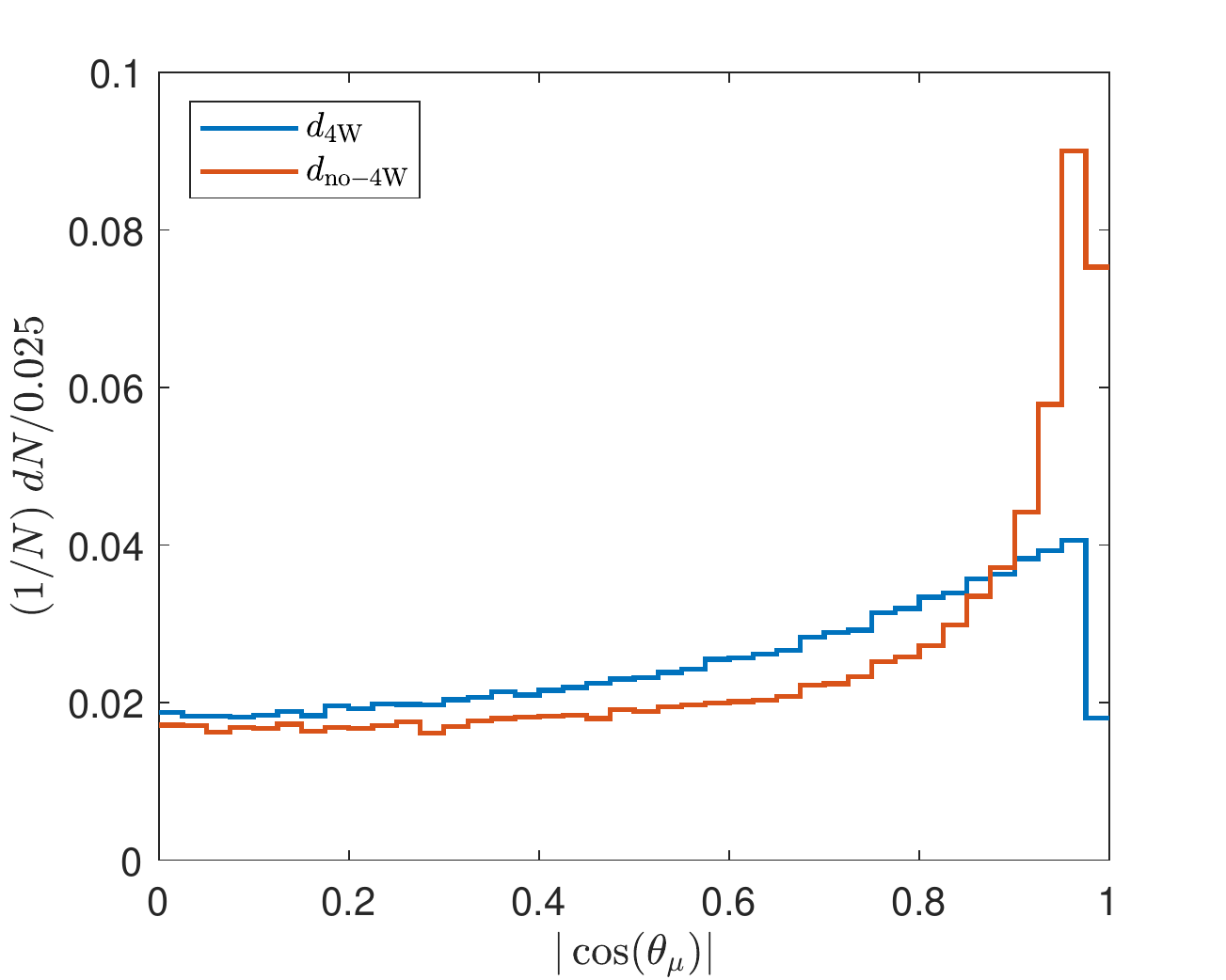}
\caption{\label{fig:vbsfeature}The normalized distributions of $|\cos(\theta ^{\rm miss})|$~(left panel) and $|\cos (\theta _{\mu})|$~(right panel).}
\end{center}
\end{figure}

The mechanism behind the ANN is a ``black box''. Thus, before using the ANN, we must verify that the data set contains information that allows us to extract the $W^+W^-\to W^+W^-$ contribution.

For the process $\mu^+\mu^-\to \nu_{\mu}\bar{\nu}_{\mu}W^+W^-$, the (anti-)neutrinos tend to be along the muon beam direction and are back-to-back.
Meanwhile, when $\hat{s}$ is large, $W^{\pm}$ are energetic and also back-to-back.
Consequently, the (anti-)neutrinos from $W\to \ell \nu$ tend to be along the directions of $W^{\pm}$ and therefore also back-to-back.
One can conclude that, the transverse missing momentum in the $W^+W^-\to W^+W^-$ contribution should be relatively small.
At a lepton collider, all the components of the missing momentum can be obtained by using momentum conservation with satisfactory accuracy, and the zenith angle of the missing momentum~(denoted as $\theta ^{\rm miss}$) can be regarded as an observable.
We find that $\theta ^{\rm miss}$ provides a better discrimination than the transverse missing momentum.
At $\sqrt{s}=30\;{\rm TeV}$ the normalized distributions of $|\cos(\theta^{\rm miss})|$ for $O_{M_0}$ are shown in left panel of Fig.~\ref{fig:vbsfeature}.
One can see that $|\cos(\theta^{\rm miss})|$ is indeed closer to $1$ for events from $d _{\rm 4W}$.

The cross-section can also be contributed by other VBS processes in addition to $W^+W^-\to W^+W^-$.
Taking the process $\mu^+\mu^-\to \nu\nu\bar{\nu}\bar{\nu}e^{\pm}\mu^{\mp}$ as an example, there is also $ZW^{\pm}\to ZW^{\pm}$ process as shown in the second Feynman diagram in Fig.~\ref{fig:diagram}~(b).
For the $ZW^{\pm}\to ZW^{\pm}$ process, the direction of the muon in the final state tends to be along the direction of ${\bf z}$-axis.
Denoting the zenith angle of the muon as $\theta _{\mu}$, the normalized distributions of $|\cos(\theta_{\mu})|$ for $O_{T_0}$ are shown in the right panel of Fig.~\ref{fig:vbsfeature}.
One can see that $|\cos(\theta_{\mu})|$ is closer to $1$ for events from $d _{\rm no-4W}$.

Although we can distinguish events from $d _{\rm 4W}$ and $d _{\rm no-4W}$ by some observables in this way, the distinction is not very efficient. One has to analyze different processes accordingly.
To achieve a desirable efficiency, a complicated analysis is required.
Finding patterns from complicated relationship is what an ANN is good at.
It can be seen that even without the ANN, there are still clues to distinguish events from $d _{\rm 4W}$ and $d _{\rm no-4W}$.
The ANN simply automates and improves the search for the clues.

%%%%%%%%%%%%%%%%%%%%%%%%%%%%%%%%%%%%%%
\subsection{\label{sec3.2}The neural network approach}
%%%%%%%%%%%%%%%%%%%%%%%%%%%%%%%%%%%%%

\begin{figure}
\begin{center}
\includegraphics[width=0.8\textwidth]{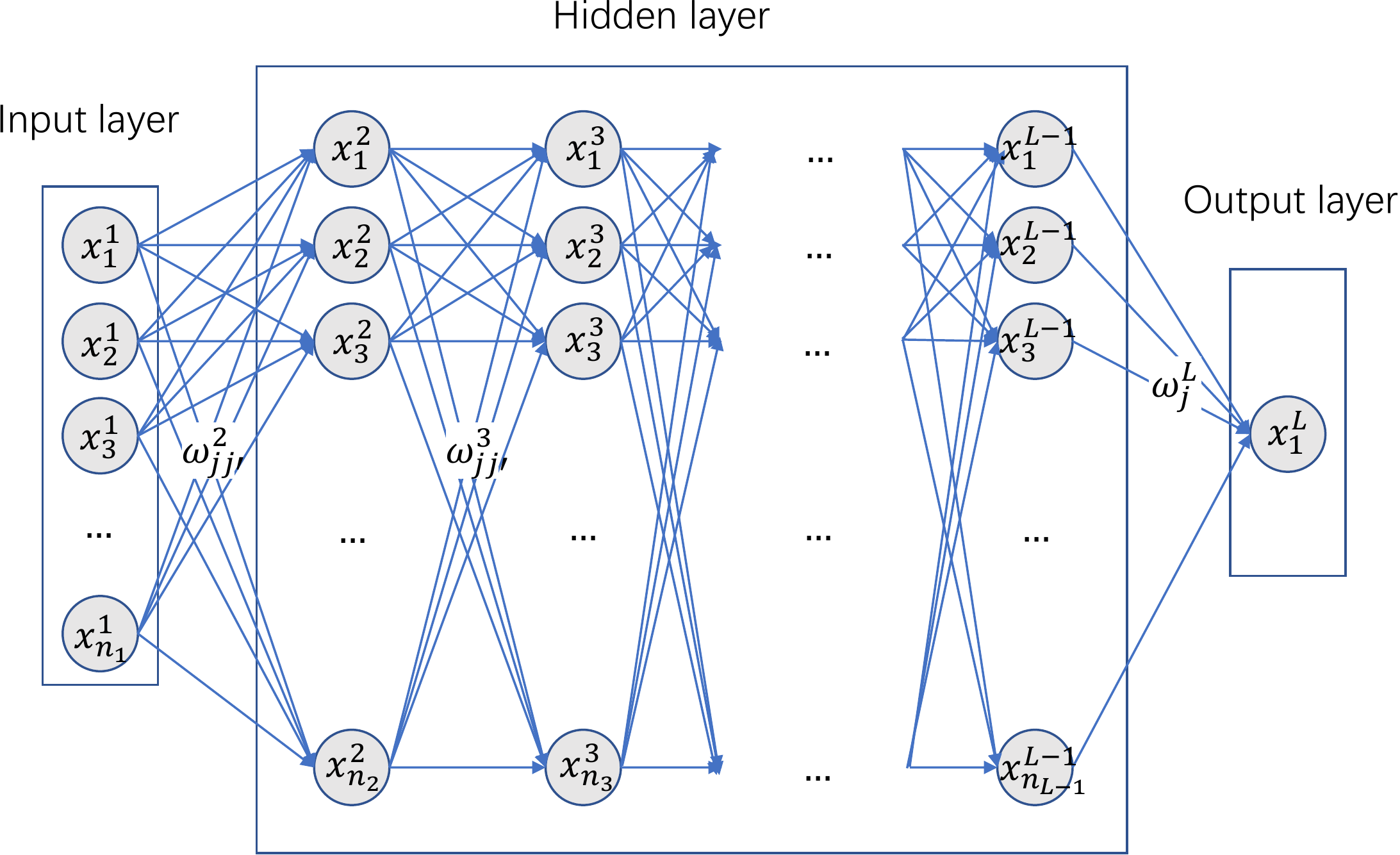}
\caption{\label{fig:ann}The graphical representation of the ANN.}
\end{center}
\end{figure}

The ANN is a mathematical model to simulate a human brain~\cite{ann}, which is good at finding the complicated mathematical mapping relationship between input and output. It could be utilized for clustering the events.
However, a classification is not tunable. For example, to archive a cleaner event set of $d _{\rm 4W}$, some events from the $d _{\rm 4W}$ could be allowed to be misidentified as from $d _{\rm no-4W}$.
Therefore, instead of clustering, we treat the identification of the events from $d _{\rm 4W}$ as a regression problem.

For an ANN, the relationship between input and output is determined by interconnected nodes and their connection modes.
We use a dense connected ANN.
An ANN is composed with an input layer, hidden layers and an output layer.
Denoting $x_j^i$ as neurons in the $i$-th layer, where $x_{1\leq j \leq n_1}^1$ are input neurons, $x_{1 \leq j \leq n_i}^{2\leq i\leq L-1}$ are in hidden layers and $x_1^L$ is the output neuron, where $L$ is the number of layers and $n_i$ are the number of neurons in the $i$-th layer, the ANN can be depicted in Fig.~\ref{fig:ann}.

A value is assigned for each neuron which is also denoted as $x_j^i$.
Then $x_{j'}^{i+1}$ can be related with $x_j^i$ as
\begin{equation}
x_{j'}^{i+1}=f_{j'}^{i+1}\left(\sum _j \omega _{jj'}^{i+1}x_j^i + b_{j'}^{i+1}\right),
\label{eq.3.3}
\end{equation}
where $\omega _{jj'}^{i+1}$ and $b_{j'}^{i+1}$ are trainable parameters.
$\omega _{jj'}^{i+1}$ are called the elements of the weight matrix $W^{i+1}$, $b_{j'}^{i+1}$ are components of the bias vector, and $f_{j'}^{i+1}$ are activation functions.
Except for the output layer, the activation functions are chosen as the parametric rectified linear unit~(PReLU) function~\cite{prelu} defined as
\begin{equation}
\begin{split}
&f(x)=\left\{\begin{array}{cc}x, & x\geq 0 ;\\ \alpha x, & x<0, \end{array}\right.\\
\end{split}
\label{eq.3.4}
\end{equation}
where $\alpha$'s are also trainable parameters.
For the output layer, no activation function~(i.e., linear activation function) is used.
We use $L=15$, $n_{10>i>1}=50$, $n_1=14$ the same as the dimension of input data and $n_L=1$ for the output layer.
The architecture is built using \verb"Keras" with a \verb"TensorFlow"~\cite{tensorflow} backend. 
The data preparation is performed by \verb"MLAnalysis"~\cite{Guo:2023nfu}.

The training and validation data-sets are prepared by using M.C. simulation.
The data-sets consist of elements with $15$ variables.
The first $14$ variables make up a $14$-dimension vector fed to the input layer~(denoted as $v_i$), and the last variable corresponds to the output layer.
$12$ components of the $14$-dimension vector are the components of 4-momenta of charged leptons, and those of the missing momentum.
The other $2$ components correspond to the flavors of the charged leptons.
The variable for output is set to $1$ if the event is from $\sigma _{\rm 4W}$ otherwise is set to $0$.
The ground truth of the output is determined with the help of non-observables, and the goal of the ANN is to reproduce the ground truth with only observables.
In the following, the output predicted by the ANN is denoted as $W_{\rm score}$. Therefore, if the ANN is well-trained, we expect that $W_{\rm score}$ is close to $1$ for the events from $\sigma _{\rm 4W}$ otherwise is close to $0$.
Two ANNs are trained, one for $O_{M_i}$ operators and the other for $O_{T_i}$ operators.
For each operator of $O_{M_i}$ and $O_{T_i}$, one million events are generated. One half of them form the training data-sets, and the other half form the validation data-sets.
The data-sets are normalized using the z-score standardization, i.e., $v'_i$ instead of $v_i$ which is defined as $v'_i=(v_i-\bar{v_i})/\sigma _{v_i}$, where $\bar{v_i}$ and $\sigma _{v_i}$ are the mean value and the standard deviation of all $i$-th variables of the elements in the training data-sets.

\begin{figure}
\begin{center}
\includegraphics[width=0.48\textwidth]{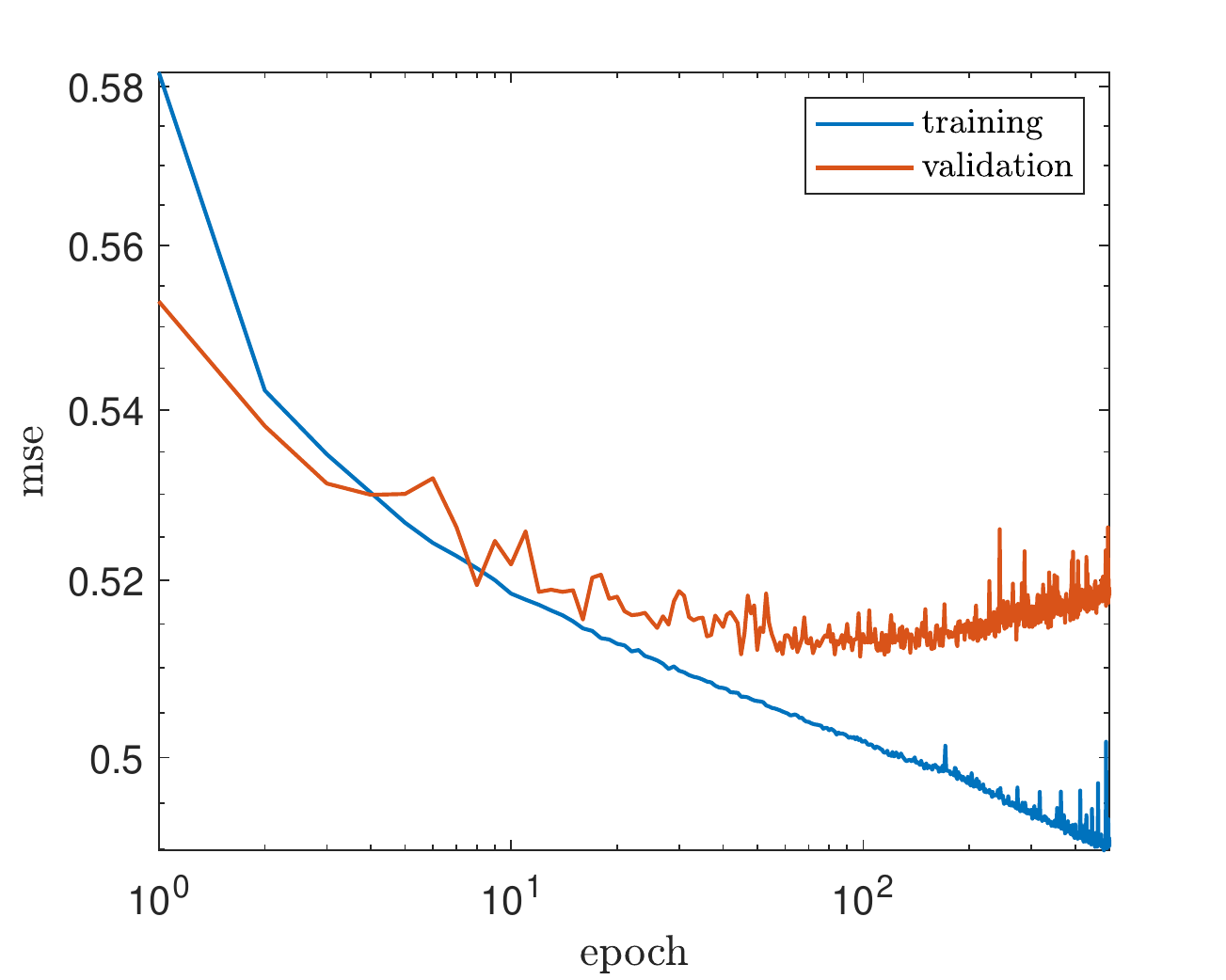}
\includegraphics[width=0.48\textwidth]{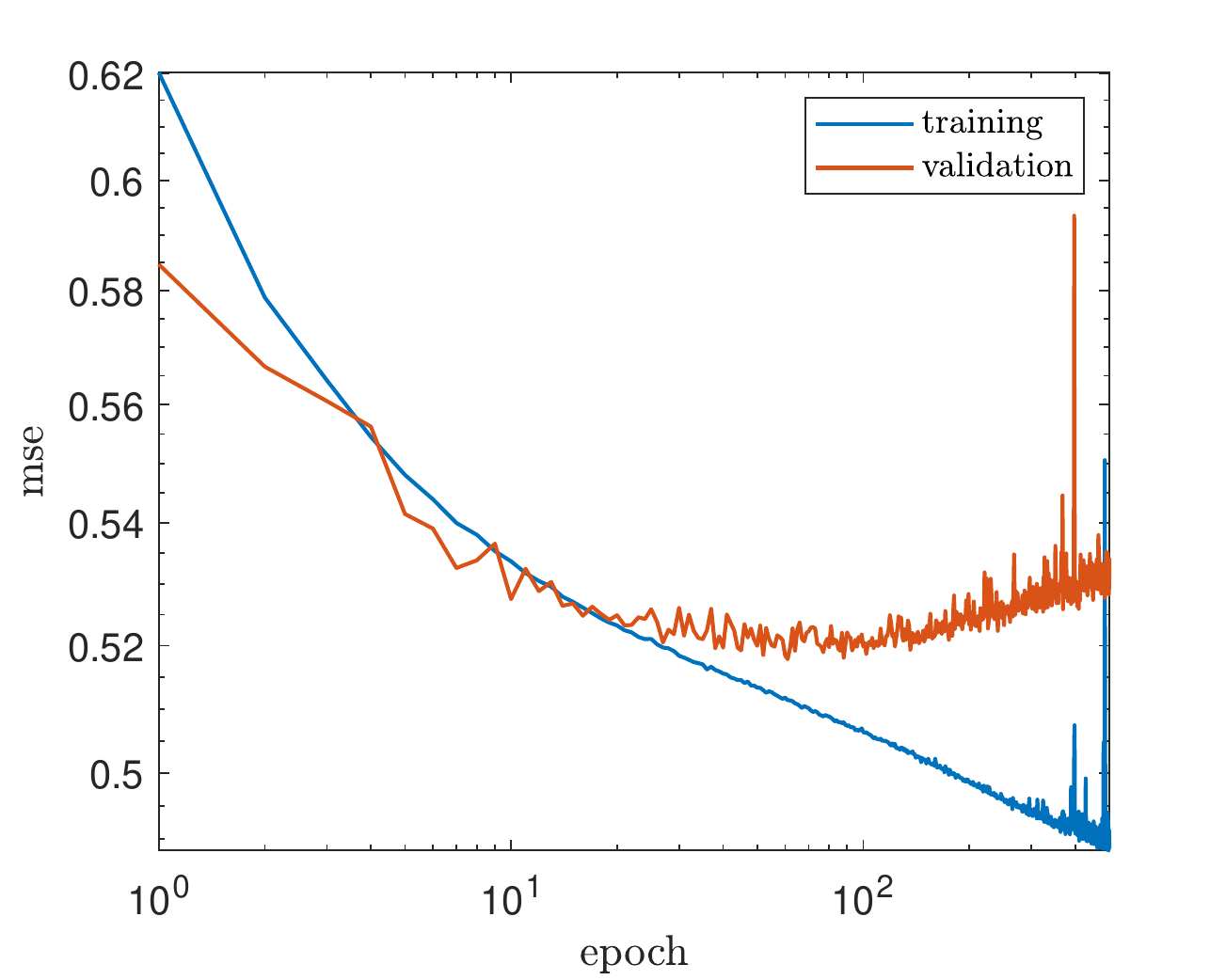}
\caption{\label{fig:learningcurvetag}The learning curves of the ANNs trained for $W_{\rm score}$, corresponding to $O_{M_i}$ (left panel) and $O_{T_i}$ (right panel).
%The left panel corresponds to $O_{M_i}$, the right panel corresponds to $O_{T_i}$.
}
\end{center}
\end{figure}

\begin{figure}
\begin{center}
\includegraphics[width=0.48\textwidth]{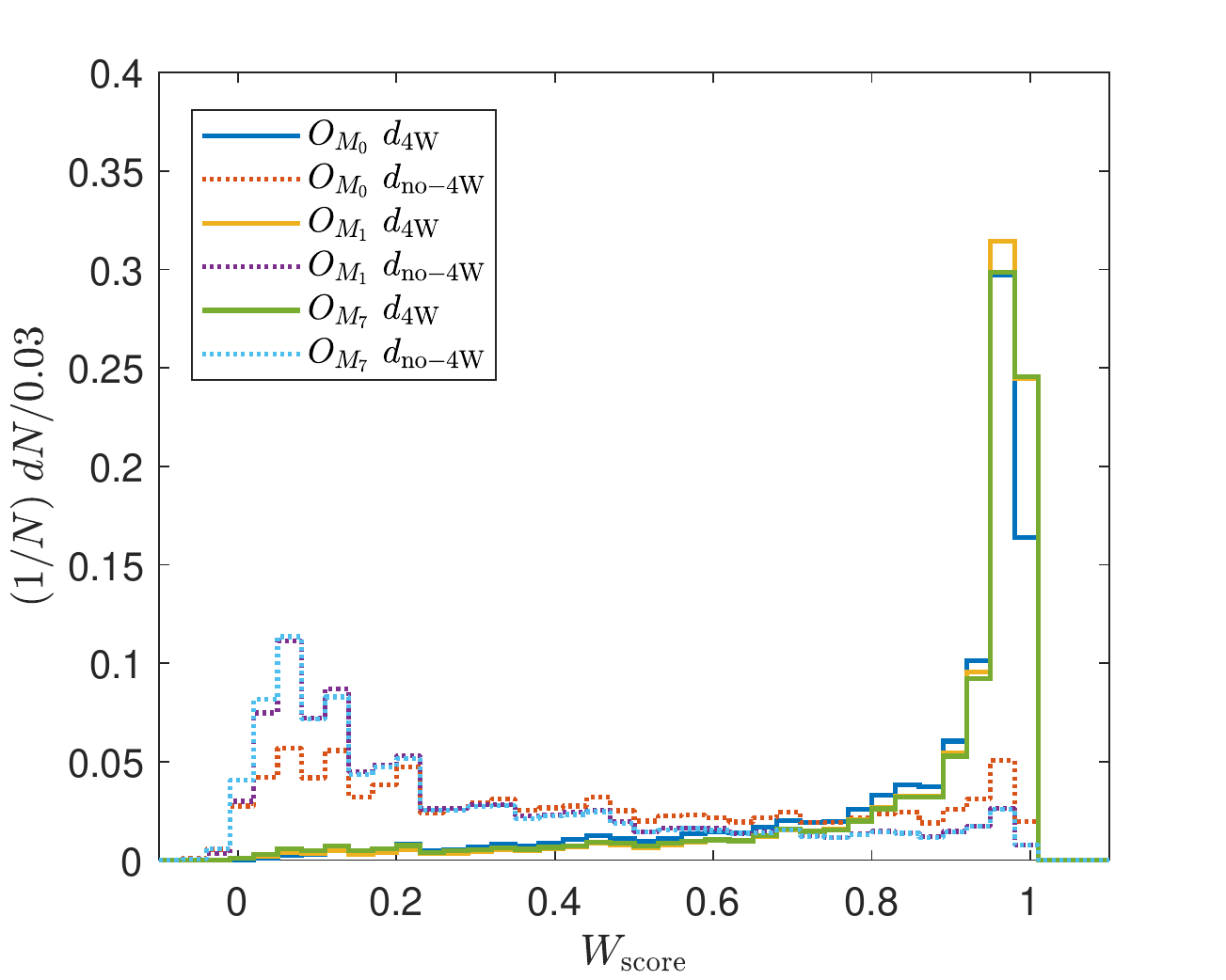}
\includegraphics[width=0.48\textwidth]{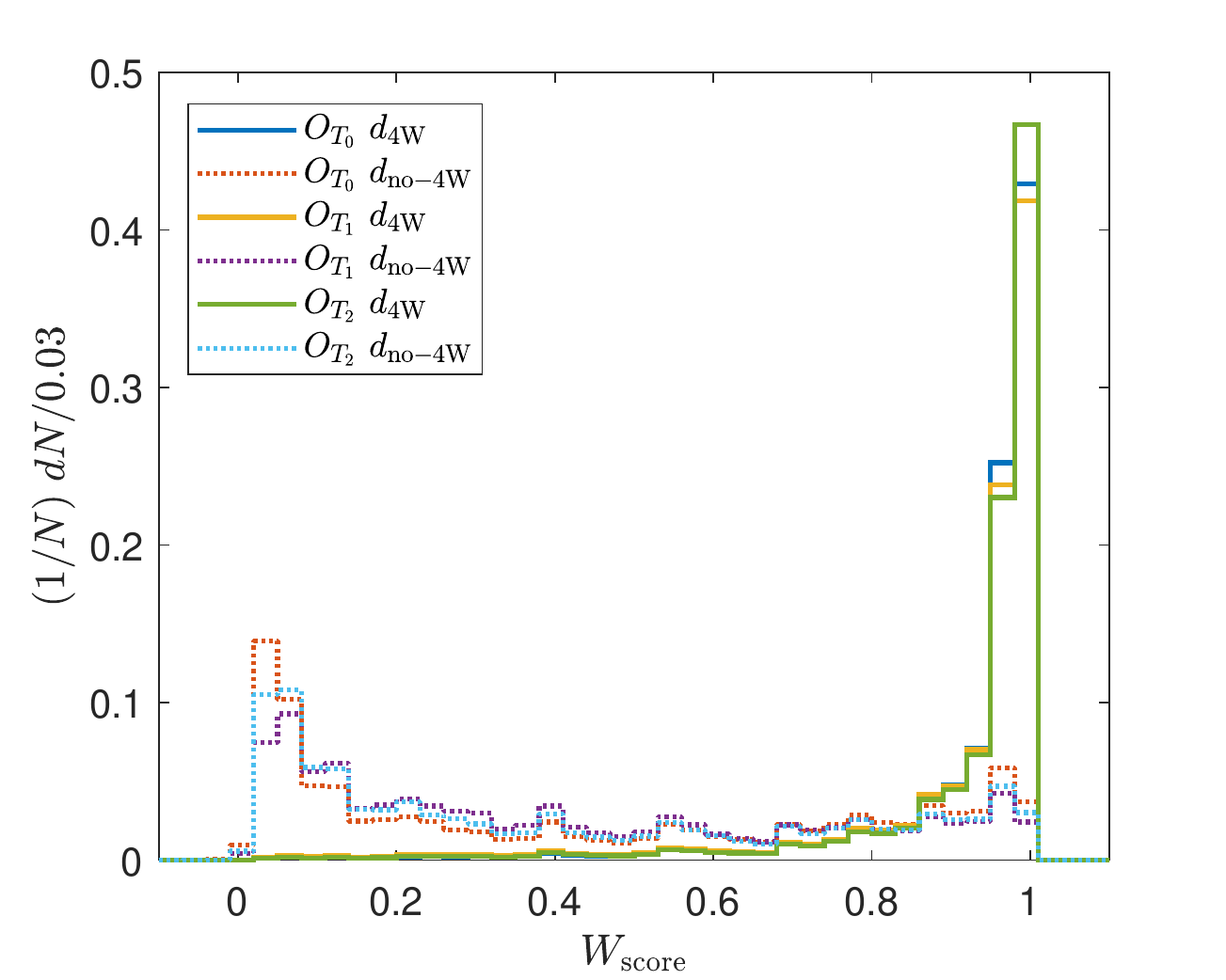}
\caption{\label{fig:wscore}The normalized distributions of $W_{\rm score}$.}
\end{center}
\end{figure}

The learning curves are shown in Fig.~\ref{fig:learningcurvetag}.
One can see that the mean squared errors~(mses) stop to decrease for the validation data-sets at about ${\rm epoches}=100\sim 200$.
To avoid overfitting, we stop at ${\rm epoches}=100$ where the mses of validation data-sets stop to decrease.
After training, the $W_{\rm score}$ results for the validation data-sets are shown in Fig.~\ref{fig:wscore}.
Compared with Fig.~\ref{fig:vbsfeature}, the $W_{\rm score}$ has stronger discrimination power.
In the following, for each operator, we choose one minimal cut of $W_{\rm score}$ as long as the mistag rate of the events from $d _{\rm 4W}$ reaches about $5\%$.
The cuts and the effects of the cuts are shown in Table~\ref{tab.fraqtionwscore}.

\begin{table}
\begin{center}
\begin{tabular}{c|c|c|c|c}
           & before cut   & $W_{\rm score}>0.85$ & $W_{\rm score}>0.9$  & $W_{\rm score}>0.95$ \\
\hline
 $O_{M_0}$ & $96.4:21.8$  & $64.6:3.3$           &                      &                      \\
 $O_{M_1}$ & $132.6:64.5$ & $99.5:5.2$           &                      &                      \\
 $O_{M_7}$ & $98.5:74.2$  &                      &                      &  $53.6:2.5$          \\
 $O_{T_0}$ & $149.0:39.5$ &                      & $117.4:5.8$          &                      \\
 $O_{T_1}$ & $141.5:36.4$ & $116.7:5.4$          &                      &                      \\
 $O_{T_2}$ & $223.0:42.0$ & $190.8:7.0$          &                      &                      \\
\end{tabular}
\end{center}
\caption{\label{tab.fraqtionwscore}$\sigma _{\rm 4W}:\sigma _{\rm no-4W}\;(\rm pb)$ for different operators after $W_{\rm score}$ cuts.}
\end{table}

%%%%%%%%%%%%%%%%%%%%%%%%%%%%%%%%%%%%
\section{\label{sec4}The reconstruction of center of mass energy of the \texorpdfstring{$W^+W^-\to W^+W^-$}{W+W- to W+W-} subprocess}
%%%%%%%%%%%%%%%%%%%%%%%%%%%%%%%%%%%%

The $\hat{s}$ of the process $W^+W^-\to W^+W^-$ is important in the study of the SMEFT, because as an EFT the Wilson coefficients should be dependent in energies.
On the other hand, the SMEFT is only valid below certain energy scale. The violation of unitarity is often used as a signal that SMEFT is no long valid, and unitarity bounds depend on the energy scale.
In any case, energy scale is important information in the study of the SMEFT.
However, the process $\mu^+\mu^-\to \nu\nu\bar{\nu}\bar{\nu}\ell^+\ell^-$ have four (anti-)neutrinos in the final state, which causes problems to reconstruct $\hat{s}$.

%\TL{xxxxxxxxxxxxxxxxxx}

\subsection{\label{sec4.1}The traditional approach}

In traditional approach, one has to analyze the kinematics of the process.
%\TL{For example, a naive approximation is to use the average. (??????)}
%
With four (anti-)neutrinos, the kinematic feature is difficult to analyze.
%Here we briefly introduce an example.
For the process $\mu^+\mu^-\to \nu\bar{\nu}W^+W^-$, the (anti-)neutrinos tend to be along the beam direction of $\mu^+\mu^-$ and thus the transverse momenta of the (anti-)neutrinos are small.
If one neglects the transverse momenta of (anti-)neutrinos along the beam, the transverse missing momentum is the sum of the transverse momenta of the (anti-)neutrinos in the processes $W\to \ell \nu$.
Then, further assuming the mass of the $W$ boson is negligible compared with $\hat{s}$, the situation is approximately the same as that investigated in Refs.~\cite{wwwwunitary,aaww} which deal with the reconstruction of $\hat{s}$ with two (anti-)neutrinos, and $\hat{s}$ can be approximately given by
\begin{equation}
\begin{split}
&\sqrt{\hat{s}_{\rm lep}}\approx c_1+c_2(E_{\ell^+}+E_{\ell ^-})+\left(c_3+c_4(E_{\ell^+}+E_{\ell ^-})+c_5E_{\ell^+}E_{\ell^-}\right)\cos (\theta _{\ell \ell}),
\end{split}
\label{eq.4.1}
\end{equation}
where $E_{\ell^{\pm}}$ are the energies of charged leptons, $\theta_{\ell\ell}$ is the angle between the charged leptons, and $c_i$ are the parameters to be fitted.

Eq.~(\ref{eq.4.1}) is in fact not obtained by kinematic analysis but by using machine learning approach.
With the help of some kinematic analysis and approximations, we translate our problem with four (anti-)neutrinos into the one with two (anti-)neutrinos solved by Eq.~(\ref{eq.4.1}).
Using kinematic analysis, there is another approximation which is $\hat{s}_{\rm ap}$ in Refs.~\cite{wwwwunitary,aaww}.
$\hat{s}_{\rm ap}$ is less accurate than $\hat{s}_{\rm lep}$, therefore is not discussed in this paper.
However, the procedure of deriving the $\hat{s}_{\rm ap}$ shows that, although it is not possible to give the exact $\hat{s}$ by the information in the final state, it is possible to give the most likely one.
Using the ANN is merely an improvement and an automation of this complicated procedure.

%%%%%%%%%%%%%%%%%%%%%%%%%%%%%%%%%%%%%%%%%%%%%%%%%%%%%%%
\subsection{\label{sec4.2}The neural network approach}
%%%%%%%%%%%%%%%%%%%%%%%%%%%%%%%%%%%%%%%%%%%%%%%%%%%%%

\begin{figure*}
\begin{center}
\includegraphics[width=0.32\textwidth]{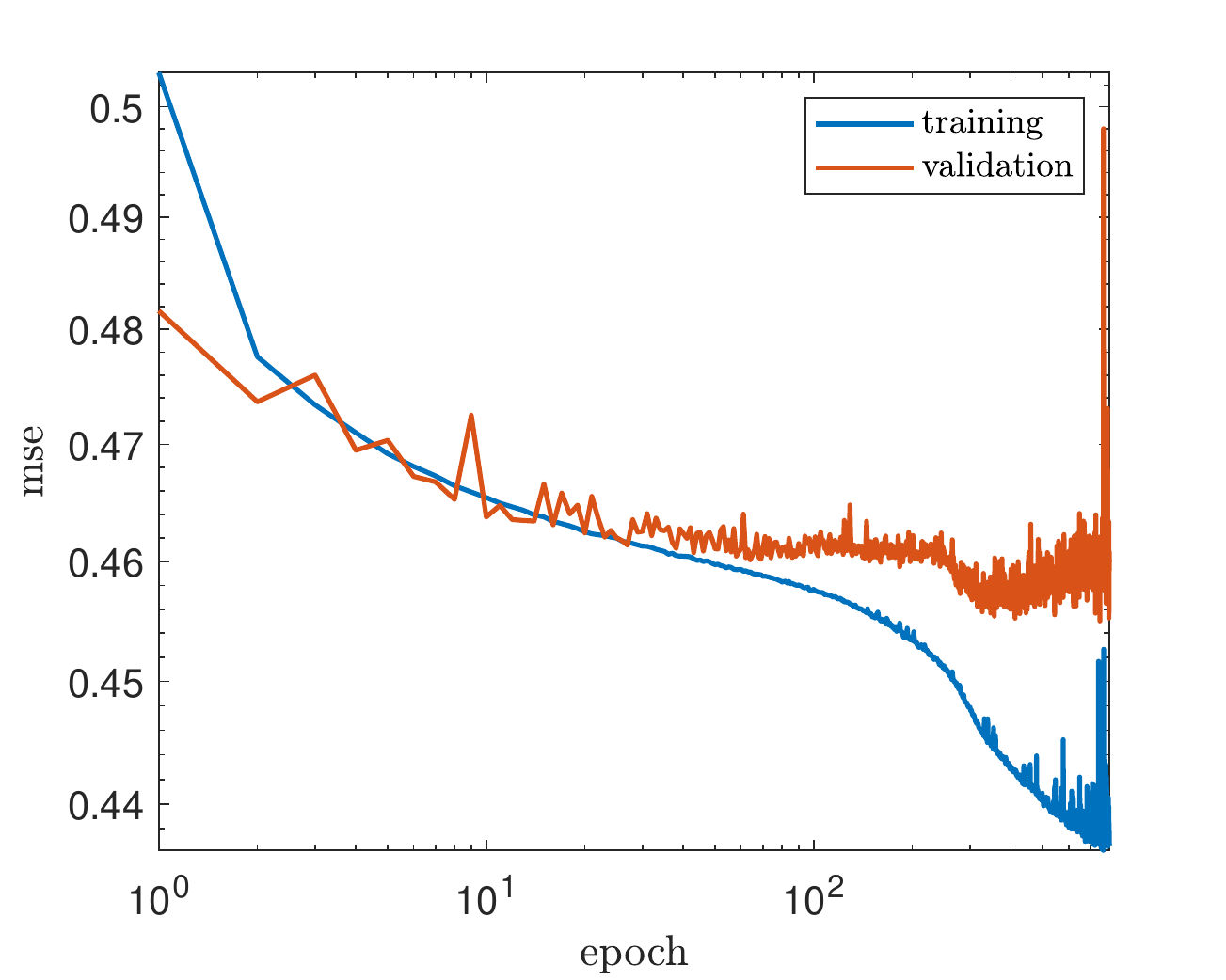}
\includegraphics[width=0.32\textwidth]{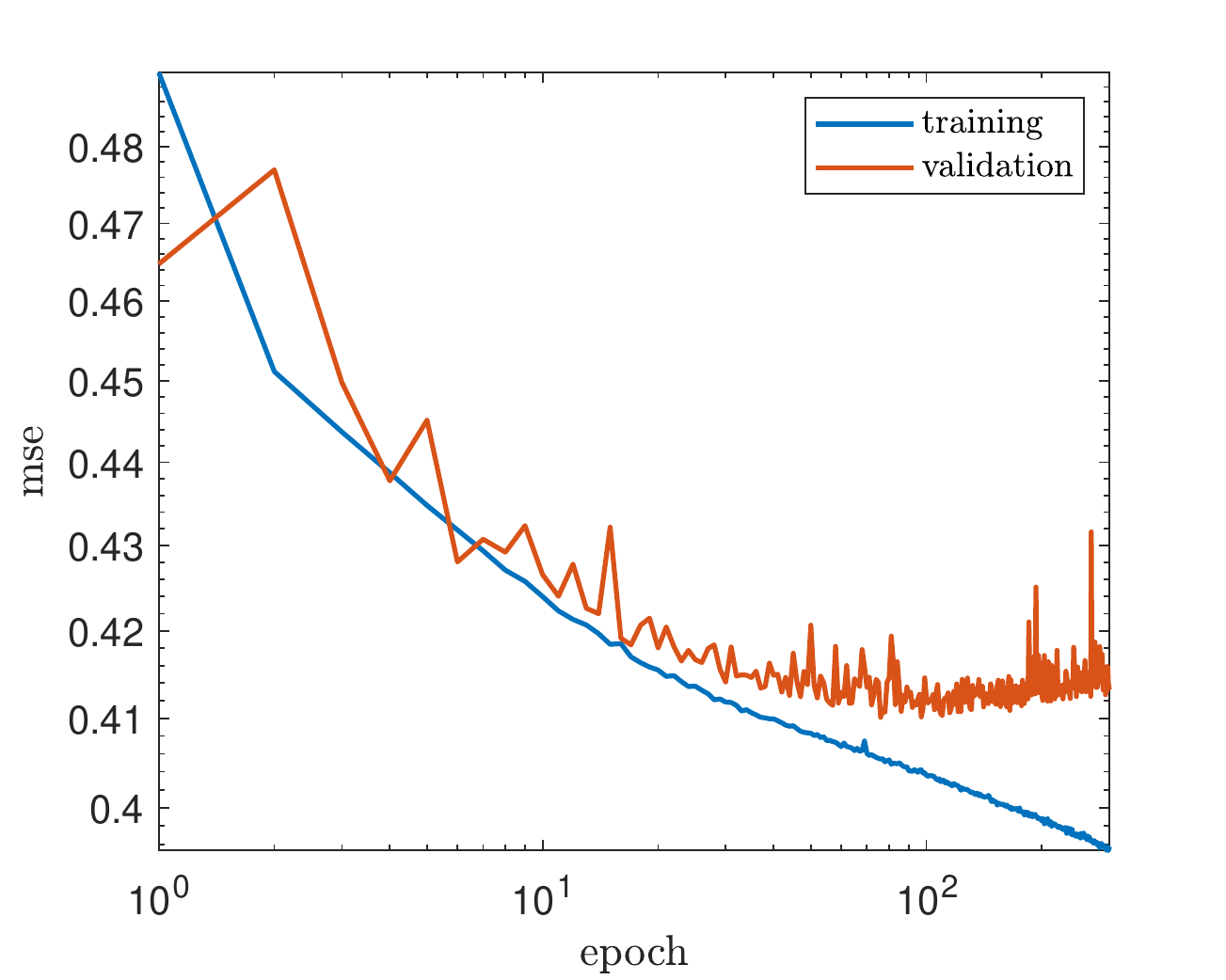}
\includegraphics[width=0.32\textwidth]{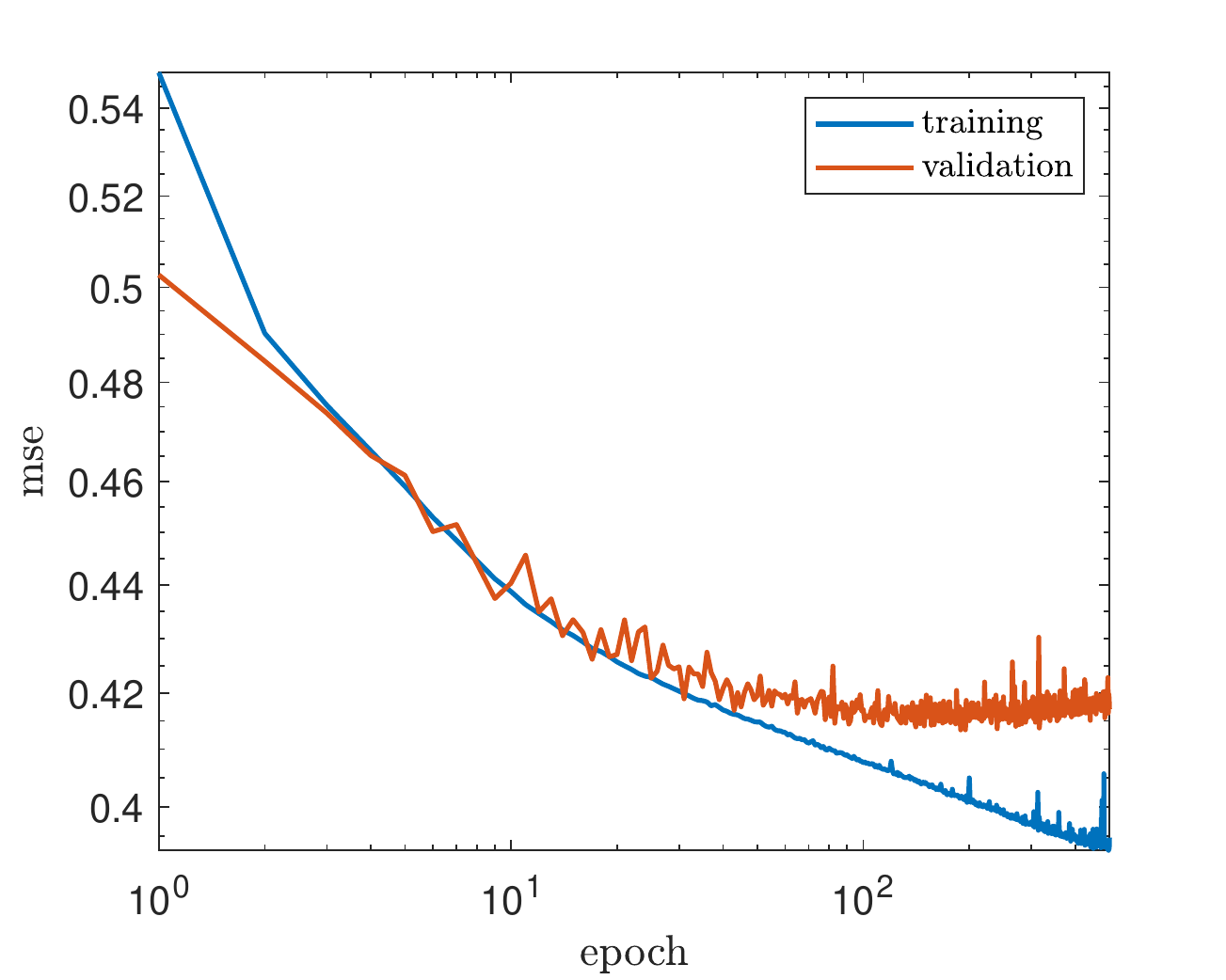}
\caption{\label{fig:learningcurveshat}The learning curves of the ANNs trained for $\sqrt{\hat{s}}$. The left (middle) [right] panel corresponds to $O_{S_i}$ ($O_{M_i}$) [$O_{T_i}$].}
\end{center}
\end{figure*}

\begin{figure*}
\begin{center}
\includegraphics[width=0.32\textwidth]{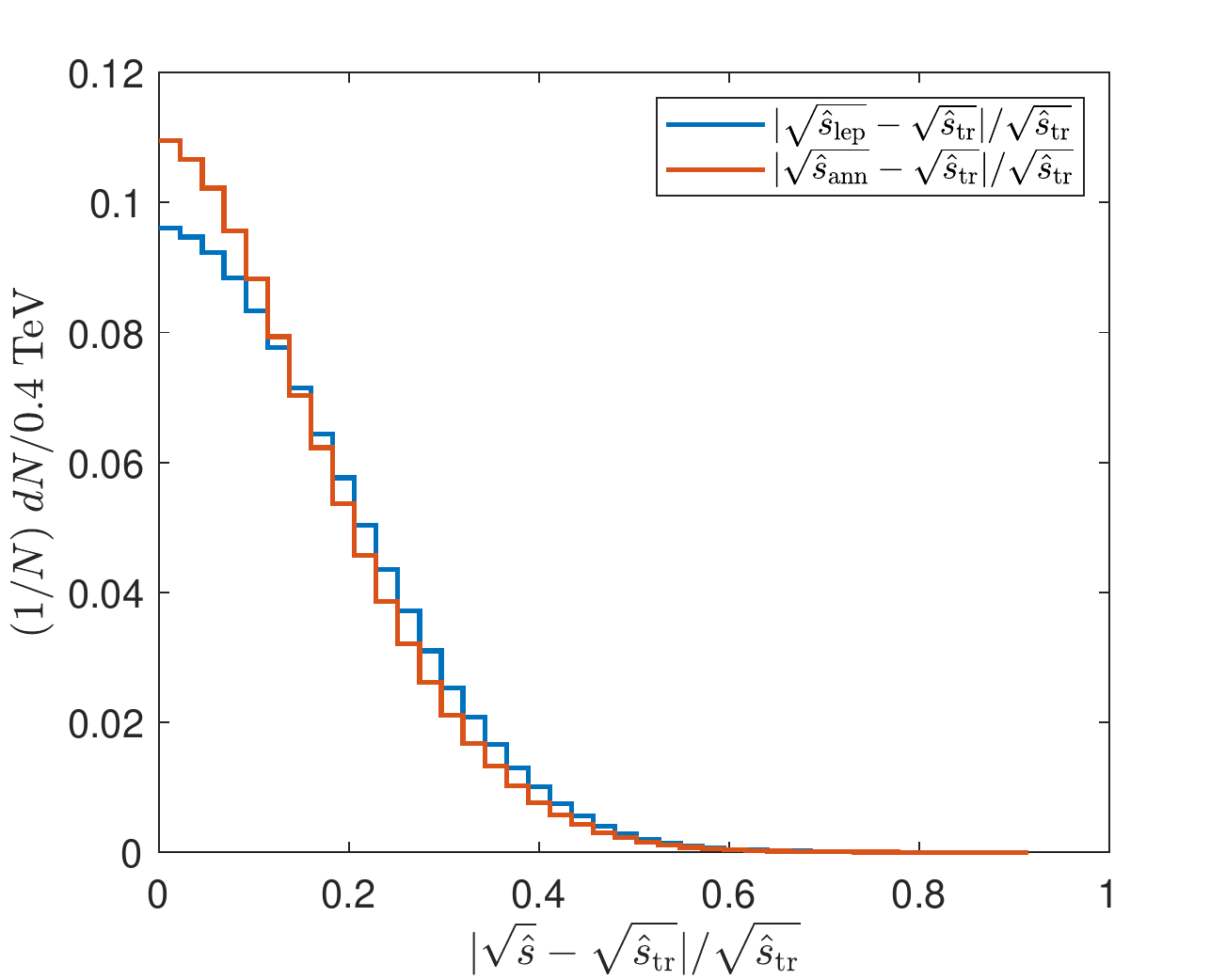}
\includegraphics[width=0.32\textwidth]{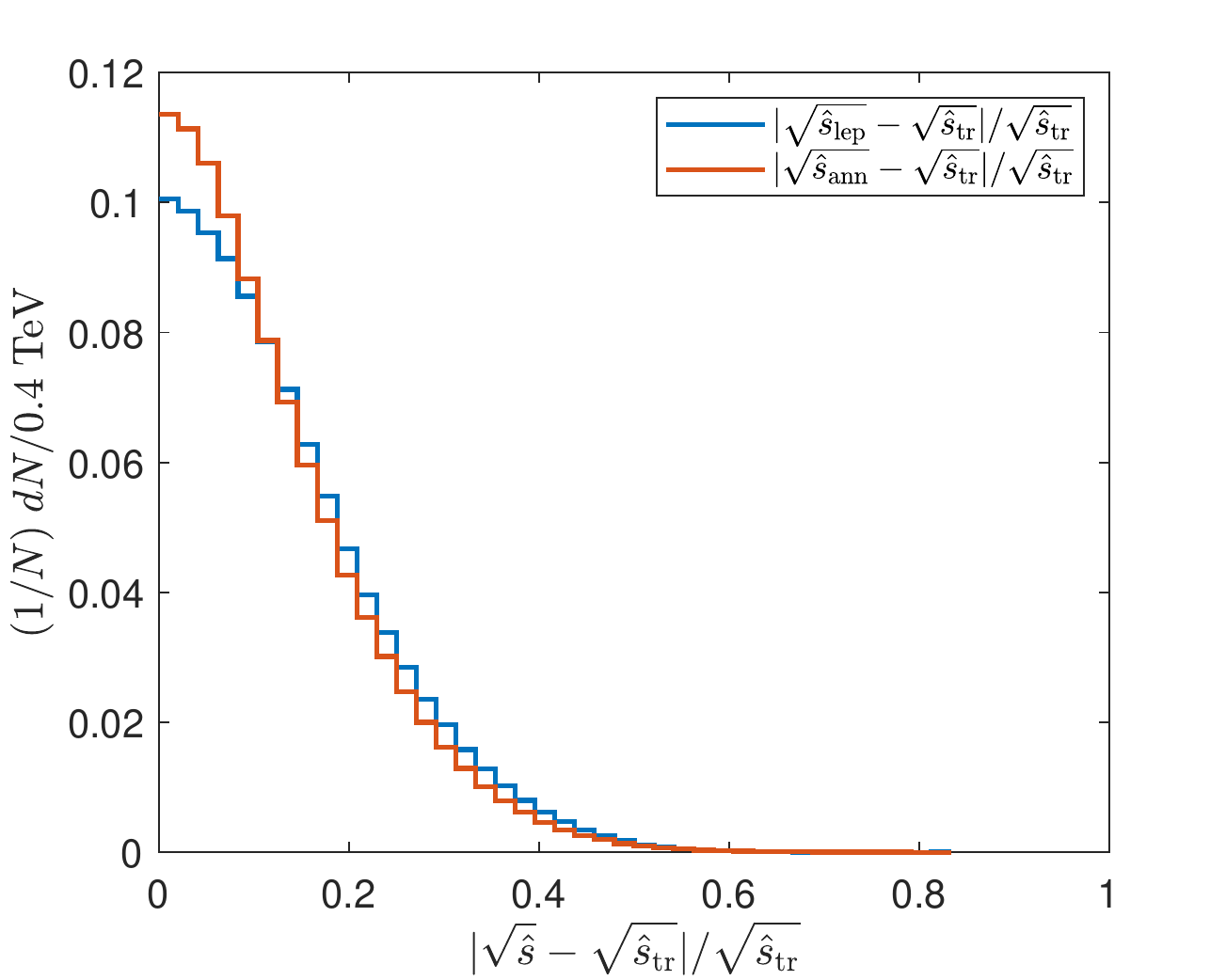}
\includegraphics[width=0.32\textwidth]{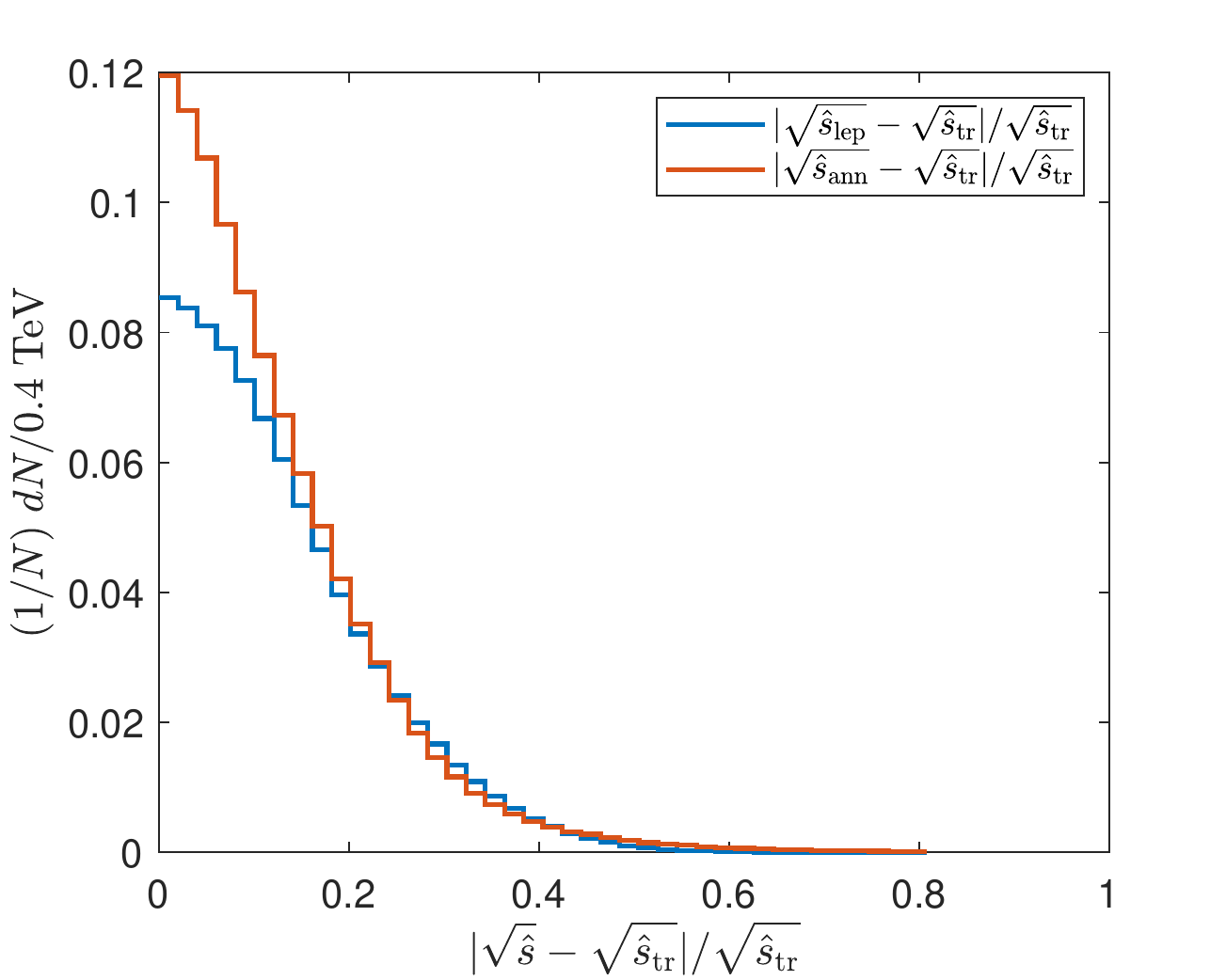}
\caption{\label{fig:shat}The normalized distributions of $|\Delta \sqrt{\hat{s}}|$. The left (middle) [right] panel corresponds to $O_{S_i}$ ($O_{M_i}$) [$O_{T_i}$].}
\end{center}
\end{figure*}

We trained three ANNs to reconstruct $\hat{s}$, which correspond to $O_{S_i}$, $O_{M_i}$ and $O_{T_i}$ operators.
The architecture of the ANNs as well as the $14$-dimensional vectors to feed the input are the same as those used in Sec.~\ref{sec3.2}.
For the output layer, the ground truth of $\hat{s}$ is estimated as $\hat{s}_{\rm tr}=(p_{\ell^+}+p_{\ell^-}+p_{\nu}+p_{\bar{\nu}})^2$, where $p_{\nu}$ ($p_{\bar{\nu}}$) is the 4-momentum of the neutrino (anti-neutrino) with the same flavor as $\ell^+$ ($\ell^-$) and with the direction closest to $\ell^+$ ($\ell^-$).
Since the Les-House event files created by \verb"MadGraph5_aMC@NLO" contains the information of intermediate $W^{\pm}$ for events from NP squared terms, it is possible to verify the correctness of $\hat{s}_{\rm tr}$.
For the events from $O_{S_0}$ and for all events containing intermediate $W^{\pm}$, we find the rate of mismatch is smaller than $0.0001\%$.

To construct the data-sets, one million events are generated for each operator.
Only events in $\sigma_{\rm 4W}$ are included, and the events are divided. One half of them are in the training data-sets, and the other half form the validation data-sets.
For $O_{S_i}$, $O_{M_i}$ and $O_{T_i}$ operators, about $1.50$, $1.03$ and $1.21$ million events are included in each data-set, respectively.
The learning curves are shown in Fig.~\ref{fig:learningcurveshat}.
To avoid overfitting, we stop at the epoch when the accuracy of validation data-set starts to fall.
Based on the learning curves in Fig.~\ref{fig:learningcurveshat}, we choose to stop at $100$, $200$ and $500$ epoches for $O_{S_i}$, $O_{M_i}$ and $O_{T_i}$ operators.
Denoting $\hat{s}_{\rm ann}$ as the prediction of $\hat{s}$ by the ANNs, for the validation data-sets, the normalized distributions of relative differences defined as $\Delta\sqrt{\hat{s}}/\sqrt{\hat{s}_{\rm tr}}$ are shown in Fig.~\ref{fig:shat}.
For comparison with $\hat{s}_{\rm lep}$, $c_i$ in Eq.~(\ref{eq.4.1}) are fitted with the training data-sets and the results are listed in Table~\ref{tab.fitted}.
The normalized distributions of relative differences for $\hat{s}_{\rm lep}$ are also shown in Fig.~\ref{fig:shat}.
For both $\hat{s}_{\rm ann}$ and $\hat{s}_{\rm lep}$, the predictions for most events can be smaller than $40\%$.
One can find that the ANNs are able to predict $\hat{s}$ more accurately than $\hat{s}_{\rm lep}$.
If one requires the relative difference to be smaller than $20\%$, using  $\hat{s}_{\rm ann}$, $76.8\%$, $81.9\%$ and $81.8\%$ of the $O_{S_i}$, $O_{M_i}$ and $O_{T_i}$ events satisfy the requirement, compared with $72.6\%$, $78.6\%$ and $66.7\%$ for using $\hat{s}_{\rm lep}$.

\begin{table}
\begin{center}
\begin{tabular}{c|c|c|c|c|c}
 & $c_1\;(\rm TeV)$ & $c_2$ & $c_3\;(\rm TeV)$ & $c_4$ & $c_5\;({\rm TeV}^{-1})$ \\
\hline
 $O_{S_i}$ & $7.39$ & $-0.131$  & $-3.64$ & $-1.11$ & $0.0667$ \\
 $O_{M_i}$ & $9.11$ & $-0.0203$ & $-5.09$ & $-0.680$ & $0.0236$ \\
 $O_{T_i}$ & $10.15$ & $-0.0448$  & $-8.94$ & $-0.141$ & $-0.0351$ \\
\end{tabular}
\end{center}
\caption{\label{tab.fitted}Results of $c_i$ in Eq.~(\ref{eq.4.1}) fitted with the training data-sets.}
\end{table}

\section{\label{sec5}Signal significance}

\subsection{\label{sec5.1}The partial wave unitarity bound}

As an EFT, the SMEFT is only valid under the NP energy scale $\Lambda$.
The large $\hat{s}$ at the muon collider provides a great chance to detect the NP. Meanwhile, the verification of the validity of the SMEFT becomes inevitable.
The partial wave unitarity has been widely used in previous studies as an indicator of the SMEFT validation~\cite{unitarity1,unitarity2,unitarity3,unitarity4,ubnew1,ubnew2,ubnew3,ntgc7,wprime}.
For a VBS process $W^+_{\lambda _1}W^-_{\lambda _2}\to W^-_{\lambda _3}W^+_{\lambda _4}$ with $\lambda _{1,2,3,4}=\pm 1, 0$ corresponding to the helicities of the vector bosons, in the c.m. frame with ${\bf z}$-axis along the flight direction of $W^-$ in the initial state, the amplitudes can be expanded as~\cite{partialwaveexpansion}
\begin{equation}
\begin{split}
&\mathcal{M}(W^-_{\lambda _1}W^+_{\lambda _2}\to W^-_{\lambda _3}W^+_{\lambda _4})=8\pi \sum _{J}\left(2J+1\right)e^{i(\lambda-\lambda ') \phi}d^J_{\lambda \lambda '}(\theta) T^J\;,\\
\end{split}
\label{eq.5.1}
\end{equation}
where $\theta$ and $\phi$ are zenith and azimuth angles of the $W^-$ boson in the final state, $\lambda = \lambda _1-\lambda _2$, $\lambda ' =\lambda _3-\lambda _4$ and $d^J_{\lambda \lambda '}(\theta)$ are the Wigner D-functions.
The partial wave unitarity bound is $|T^J|\leq 2$~\cite{partialwaveunitaritybound}.
With the helicity amplitudes calculated in Appendix~\ref{ap2}, and assuming one operator at a time, the tightest bounds are
\begin{equation}
\begin{split}
&\hat{s}^2<\frac{48\pi \Lambda^4}{|f_{S_0}|},\;\;\hat{s}^2<\frac{24\pi \Lambda^4}{|f_{S_1}|}\;\;\hat{s}^2<\frac{24\pi\Lambda^4}{|f_{S_2}|}.\\
&\hat{s}^2<\frac{32\pi\Lambda^4}{|f_{M_0}|},\;\;\hat{s}^2<\frac{128\pi \Lambda^4}{|f_{M_1}|},,\;\;\hat{s}^2<\frac{256\pi\Lambda^4}{|f_{M_7}|},\\
&\hat{s}^2<\frac{6\pi\Lambda^4}{|f_{T_0}|},\;\;\hat{s}^2<\frac{8\pi\Lambda^4}{|f_{T_1}|},\;\;\hat{s}^2<\frac{16\pi\Lambda^4}{|f_{T_2}|}.\\
\end{split}
\label{eq.5.4}
\end{equation}

In this paper, the unitarity bounds are applied using a matching procedure~\cite{matchingidea2,matchingidea3} which has been used in previous studies of aQGCs at the LHC~\cite{wastudy,zastudy,wwwwunitary}.
We compare the cross-sections with and without aQGCs under a certain energy scale.
It should be emphasized that, such unitrization procedure introduces no extra assumptions.
This is important because it has been pointed out that different unitrization methods lead to different results~\cite{unitarizationeffects}, and therefore the unitrization methods introducing extra assumptions actually break the model-independence principle of the SMEFT~\cite{vbs1}.

In fact, in our approach, no bounds or constraints are applied, despite a misleading ``bounds'' in the name of this procedure.
Studying the Wilson coefficient within an energy range is standard for an EFT because the Wilson coefficients typically depend on energy scales.
Namely, even without unitarity bounds, it is a matter of interest to compare the SMEFT and the SM within a certain energy scale.
We simply choose the energy scale according to the coefficients such that unitarity is guaranteed.

\begin{figure}
\begin{center}
\includegraphics[width=0.7\textwidth]{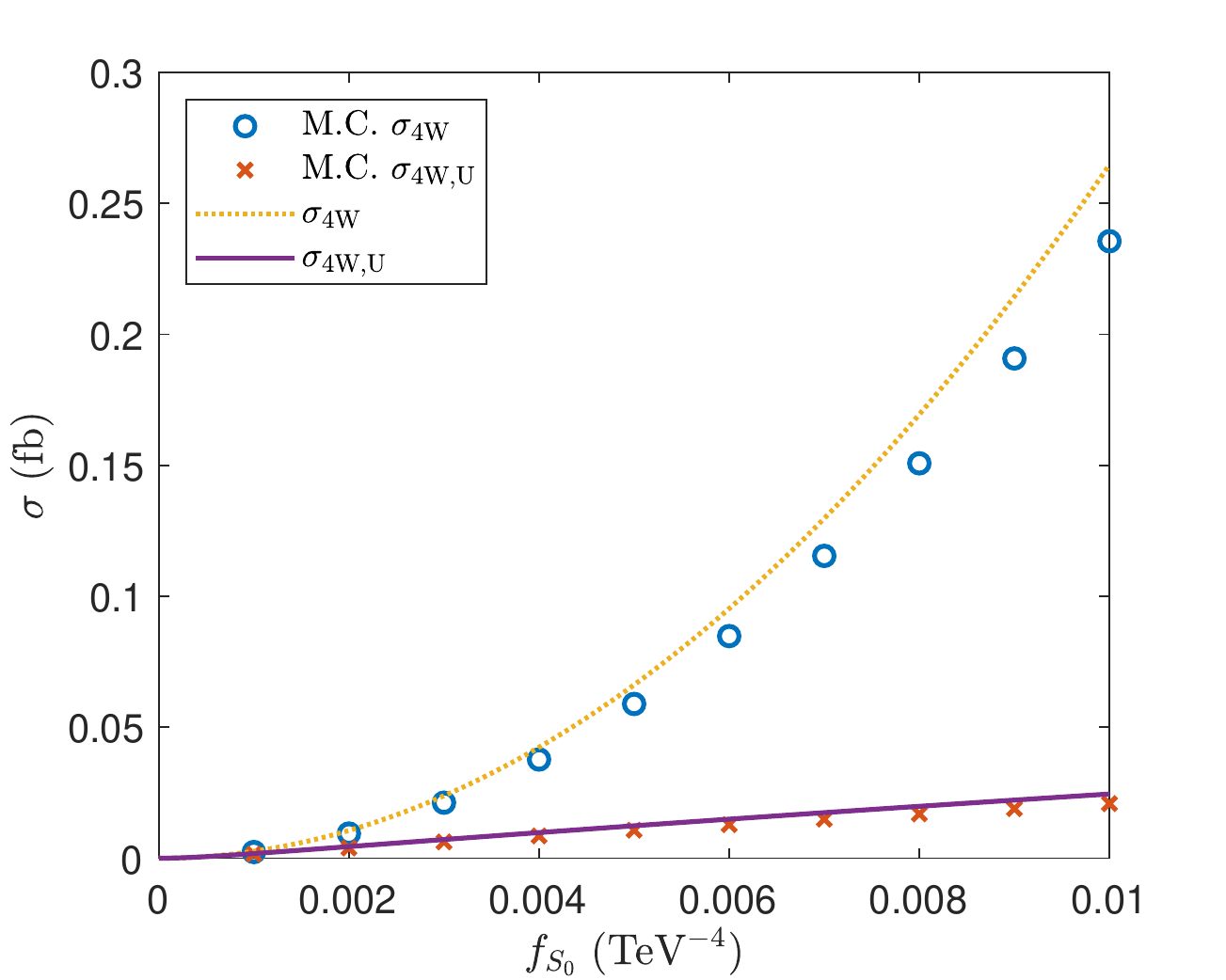}
\caption{\label{fig:effectofunitarity}$\sigma _{\rm 4W}(\mu^+\mu^-\to \bar{\nu}\nu W^+W^-\to \bar{\nu}\nu\bar{\nu}\nu\ell^+\ell^-)$ compared with $\sigma _{\rm 4W, U}(\mu^+\mu^-\to \bar{\nu}\nu W^+W^-\to \bar{\nu}\nu\bar{\nu}\nu\ell^+\ell^-)$ in Eq.~(\ref{eq.ap.6}). This is only for an illustration that the unitarity bounds are necessary.}
\end{center}
\end{figure}

The suppression of cross-section when unitarity bounds are considered has been noticed in Refs.~\cite{wastudy,zastudy,wwwwunitary}, and demonstrates the necessity of the unitarity bounds.
The effect of unitarity bounds can be estimated in terms of EVA~(see Appendix.~\ref{ap1}).
To illustrate the necessity of unitarity bounds, we compared the cases with and without unitarity bounds at $\sqrt{s}=30\;{\rm TeV}$ in Fig.~\ref{fig:effectofunitarity}.
As a verification, the results from M.C. simulation are also shown with $\sqrt{\hat{s}}_{\rm tr}$ in Sec.~\ref{sec4.2} used for unitarity bound.
It can be seen in Fig.~\ref{fig:effectofunitarity} that, the cross-section is suppressed significantly and is no longer a bilinear function of $f_{S_0}$ after the unitarity bound is applied, which is also seen in Refs.~\cite{wastudy,zastudy,wwwwunitary}.

After event selection strategy, the suppressed efficiencies of unitarity bounds will change.
Thus, we cannot use Eq.~(\ref{eq.ap.6}) to estimate the effect of unitarity bounds after the event selection strategy.
Also, we cannot use $\sqrt{\hat{s}}_{\rm tr}$ to apply the unitarity bounds in M.C. simulation because $\sqrt{\hat{s}}_{\rm tr}$ is not observable.
Instead, with ANNs trained to reconstruct $\hat{s}$ at hand, we use the $\hat{s}_{\rm ANN}$ for unitarity bounds.
Fig.~\ref{fig:effectofunitarity} is only for an illustration that the unitarity bounds are necessary.

\subsection{\label{sec5.2}Signal and backgrounds}

\begin{figure*}
\begin{center}
\includegraphics[width=0.9\textwidth]{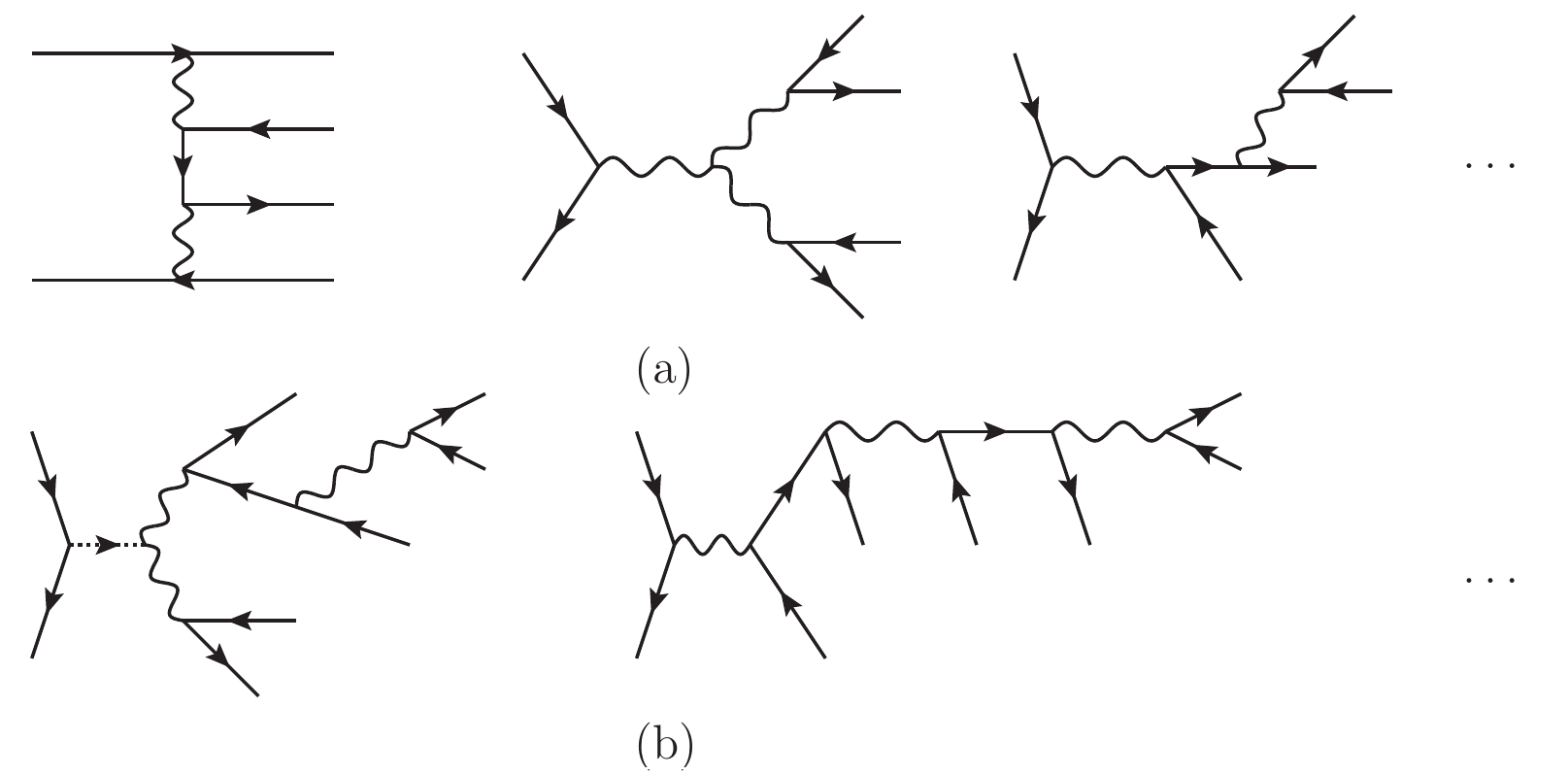}
\caption{\label{fig:diagramsm}The tree level Feynman diagrams of the SM contribution to the process $\mu^+\mu^-\to \ell^+\ell^-+\slashed{E}$.}
\end{center}
\end{figure*}
%\begin{table}
%\begin{center}
%\begin{tabular}{c|c|c|c|c|c|c|c|c|c}
%& $O_{S_0}$ & $O_{S_1}$ & $O_{S_2}$ & $O_{M_0}$ & $O_{M_1}$ & $O_{M_7}$ & $O_{T_0}$ & $O_{T_1}$ & $O_{T_2}$ \\
%\hline
%$f_X/\Lambda ^4\;(10^{-3}{\rm TeV}^{-4})$ & $17$ & $15$ & $15$ & $2$ & $4$ & $8$ & $0.2$ & $0.3$ & $0.5$ \\
%$\sigma_{\rm int}\;({\rm ab})$ & $6.33$ & $-0.94$ & $-1.2$ & $0.28$ & $18.5$ & $-19.2$ & $47.7$ & $76.5$ & $28.8$ \\
%\end{tabular}
%\end{center}
%\caption{\label{tab.range}$\sigma _{\rm 4W}:\sigma _{\rm no-4W}\;(\rm pb)$ for different operators after $W_{\rm score}$ cuts.}
%\end{table}

The major SM background is $\mu^+\mu^- \to \ell ^+ \ell ^- + \slashed{E}$.
We consider the processes with up to four (anti-)neutrinos and the Feynman diagrams are shown in Fig.~\ref{fig:diagramsm}.
Using M.C. simulation, at $\sqrt{s}=30\;{\rm TeV}$, we find the total SM cross-section as $\sigma _{\rm SM}=176.8\;{\rm fb}$.
The dominant one is $\mu^+\mu^- \to \ell ^+ \ell ^- \nu \bar{\nu}$ whose cross-section is about $155.2\;{\rm fb}$.

The signal significance is estimated using the definition $\mathcal{S}_{stat}=N_s/\sqrt{N_s+N_{bg}}$ where $N_s$ ($N_{bg}$) is the number of signal (backgrounds) events.
We define the cross-section with unitarity bounds as $\sigma _{\rm 4W,U}$ as shown in Eq.~(\ref{eq.ap.5}).
Before performing the M.C. simulation,  $\sigma _{\rm 4W,U}$ is used to initially predict the constraints on coefficients, as well as to determine the range of parameter space.
The expected luminosity of the $\mu^+\mu^-$ collider at $\sqrt{s} = 30\;{\rm TeV}$ is about $L=10\;{\rm ab}^{-1}\sim 90\;{\rm ab}^{-1}$~\cite{muoncollider5}.
By requiring $\mathcal{S}_{stat}\approx 2\sim 5$ before event selection strategies and after the unitarity bounds are applied, we choose the coefficient spaces satisfying $\sigma _{\rm 4W,U}\approx 0.1\% \times \sigma _{\rm SM}$.
The largest coefficients are calculated according to Eqs.~(\ref{eq.ap.6}-\ref{eq.ap.13}) and listed in Table~\ref{tab.cutflow}.
The coefficients for $O_{S_i}$ are much larger than $O_{M_i}$ and $O_{T_i}$, because the suppression of the unitarity bounds is more important for $O_{S_i}$ operators.
This is because the dominant helicity amplitude for each $O_{S_i}$ operator is longitudinal scattering $W_0W_0\to W_0W_0$ as shown in Eq.~(\ref{eq.ap.15}).
However, the luminosity of transverse polarized $W$ bosons from the beam is logarithmically enhanced, as shown in Eq.~(\ref{eq.ap.1}).
As a result, to produce cross-sections of the same order of magnitude, $f_{S_i}$ should be much larger than $f_{M_i}$ and $f_{T_i}$.
Meanwhile, the unitarity bounds are set by the amplitudes of $WW\to WW$, so that the suppression is at the same order of magnitude for all operators.
In return, a larger $f_{S_i}$ suffers from a more significant suppression, and an even larger $f_{S_i}$ is required as a consequence.

\begin{figure}
\begin{center}
\includegraphics[width=0.32\textwidth]{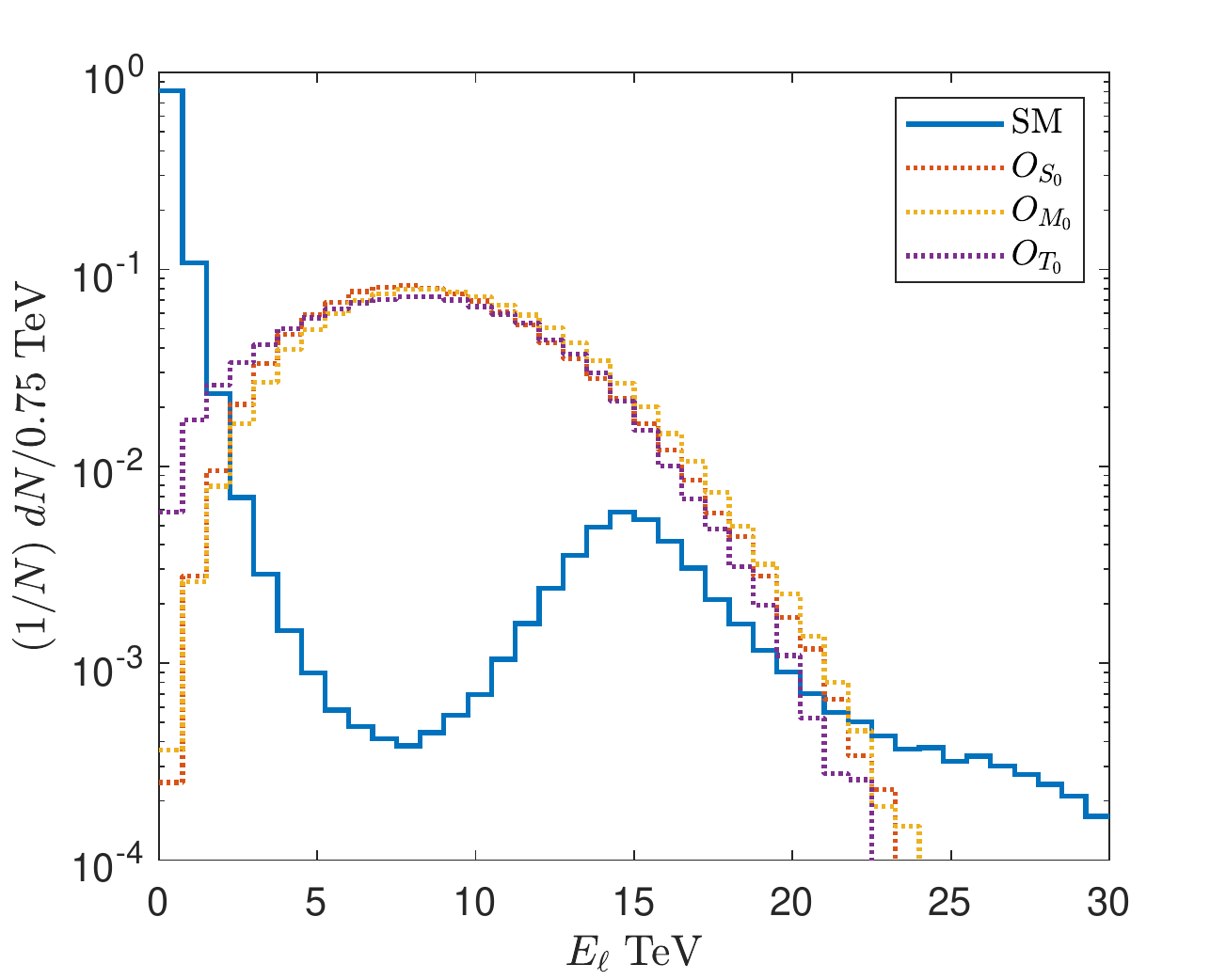}
\includegraphics[width=0.32\textwidth]{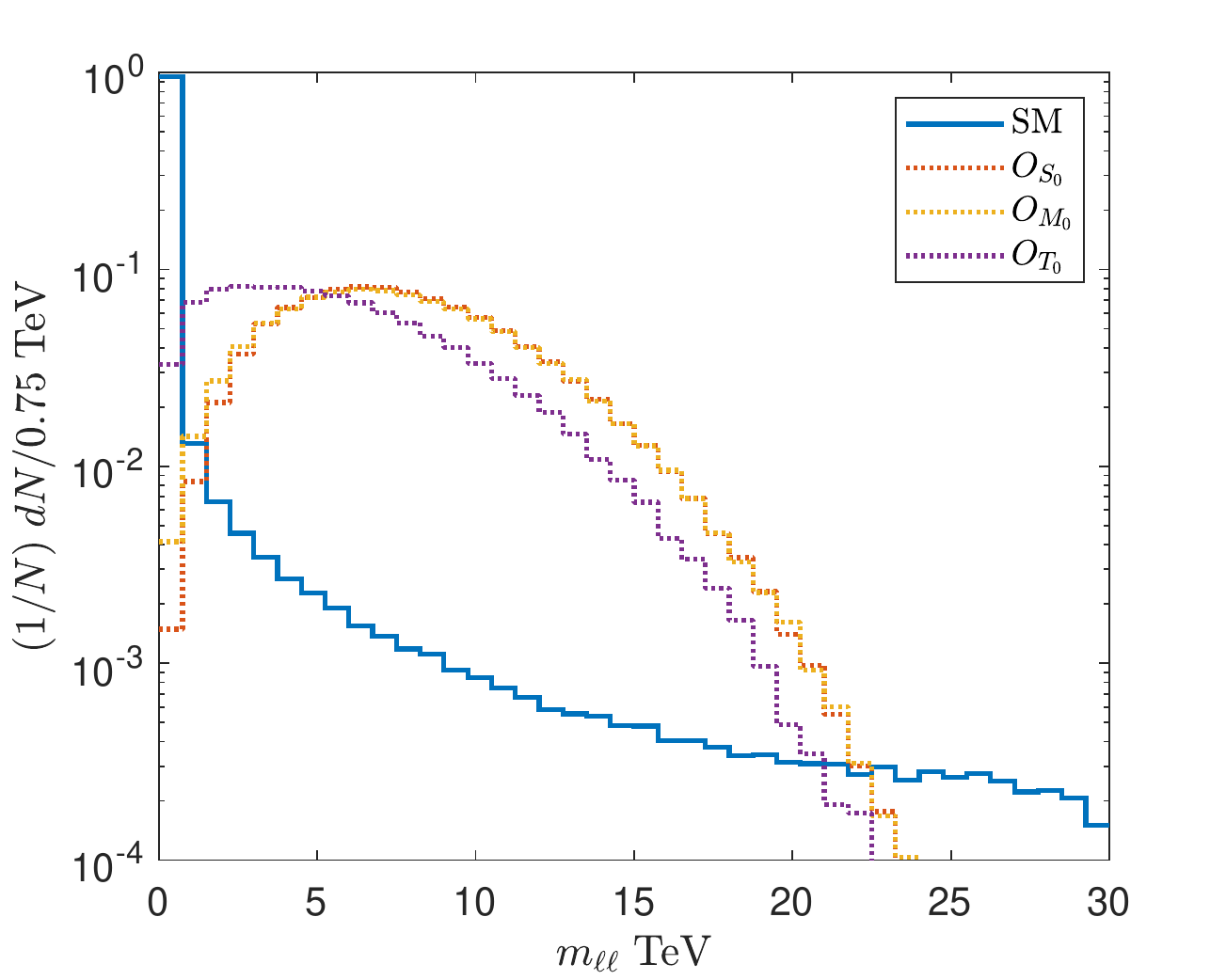}
\includegraphics[width=0.32\textwidth]{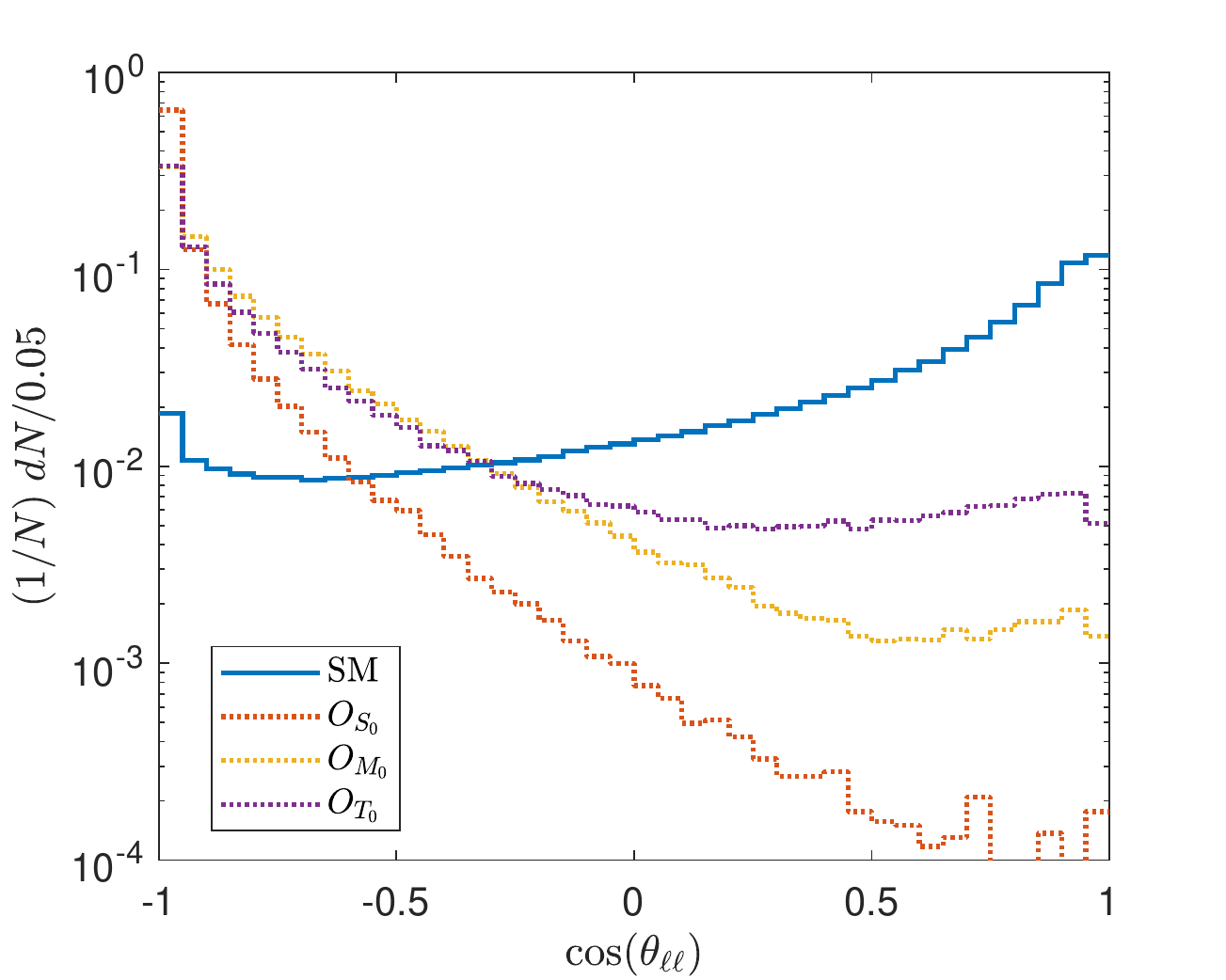}
\caption{\label{fig:kinfeature}The normalized distributions of $E_{\ell}$~(left panel), $m_{\ell\ell}$~(middle panel) and $\cos (\theta _{\ell\ell})$~(right panel).}
\end{center}
\end{figure}

To study the kinematic features of the signals and the background, in the following, a fast detector simulation is performed by \verb"Delphes"~\cite{delphes} with the muon collider card.
We cut off the events that do not contain two opposite-sign leptons or with at least two charged leptons but the hardest two have same-sign~(denoted as $N_{\ell}$ cut).
The kinematic features are shown after the $N_{\ell}$ cut.

Compared with the VBS contribution from the aQGCs, the $\hat{s}$ of the VBS in the SM is smaller.
As a result, a smaller $E_{\ell}=E_{\ell^+}+E_{\ell^-}$ is expected for the SM.
For the same reason, the invariant masses of charged leptons $m_{\ell\ell}=\sqrt{(p_{\ell^+}+p_{\ell^-})^2}$ of the signal events are larger.
For the signal events, with an energetic $W$ boson, the charged lepton should be approximately collinear to the $W$ boson.
The values of $\hat{s}$ of the signal events are large, so that the produced $W^{\pm}$ bosons tend to be approximately back-to-back for a signal event. As a result, the angle between charged leptons~(denoted as $\theta _{\ell \ell}$) should be close to $\pi$.
Taking $O_{S_0}$, $O_{M_0}$ and $O_{T_0}$ as examples, the normalized distributions of $E_{\ell}$, $m_{\ell\ell}$ and $\cos(\theta _{\ell\ell})$ are shown in Fig.~\ref{fig:kinfeature}.
It can be seen that $E_{\ell}$, $m_{\ell\ell}$ and $\cos(\theta _{\ell\ell})$ for the signal events are very different from those of the background events.

%Before the observation of the signals of the aQGCs, the goal is to constrain the coefficients.
To constrain the aQGC operators, different event selection strategies can be designed for different operators.
The event selection strategies and effects of the cuts are summarized in Table~\ref{tab.cutflow}.
We find that the event selection strategies can suppress the backgrounds significantly, while keeping most of the signal events.

\begin{table}
\begin{center}
\begin{tabular}{c|cc|c|c|c}
& \multicolumn{2}{c|}{SM} & $O_{S_0}$ & $O_{S_1}$ & $O_{S_2}$ \\
$f_{S_i}/\Lambda ^4$ & \multicolumn{2}{c|}{} & $0.1$ & $0.1$ & $0.1$ \\
\hline
$\sigma$ & \multicolumn{2}{c|}{$176805$} & $23559$ & $56280$ & $56326$ \\
$N_{\ell}$ cut & \multicolumn{2}{c|}{$140132$} & $18054$ & $43441$ & $43570$ \\
$2.2\;{\rm TeV}<E_{\ell}<22\;{\rm TeV}$ & \multicolumn{2}{c|}{$7786.7$} & $17835$ & $42912$ & $43046$ \\
$2\;{\rm TeV}<m_{\ell\ell}<20\;{\rm TeV}$ & \multicolumn{2}{c|}{$4138.8$} & $17594$ & $42331$ & $42460$ \\
$\cos(\theta_{\ell\ell})<-0.3$ & $2632.3$ & & $17378$  &         & $41758$ \\
$\cos(\theta_{\ell\ell})<-0.8$ & & $1674.8$ &          & $36049$ & \\
\hline
& \multicolumn{2}{c|}{SM}  & $O_{M_0}$ & $O_{M_1}$ & $O_{M_7}$ \\
$f_{M_i}/\Lambda ^4$ & \multicolumn{2}{c|}{} & $0.002$ & $0.004$ & $0.008$ \\
\hline
$\sigma$ & \multicolumn{2}{c|}{$176805$} & $964.0$ & $714.2$ & $955.5$ \\
$N_{\ell}$ cut & \multicolumn{2}{c|}{$140132$} & $743.8$ & $548.8$ & $738.0$ \\
$2.2\;{\rm TeV}<E_{\ell}<22\;{\rm TeV}$ & \multicolumn{2}{c|}{$7786.7$} & $735.9$ & $542.2$ & $728.5$  \\
$2.2\;{\rm TeV}<m_{\ell\ell}<20\;{\rm TeV}$ & \multicolumn{2}{c|}{$3935.0$} & $709.7$ & $516.2$ & $684.7$  \\
$W_{\rm score}>0.85$ & $964.9$ & & $402.0$ & $273.2$ & \\
$W_{\rm score}>0.95$ & & $403.8$ &         & & $220.8$ \\
\hline
& \multicolumn{2}{c|}{SM} & $O_{T_0}$ & $O_{T_1}$ & $O_{T_2}$ \\
$f_{T_i}/\Lambda ^4$ & \multicolumn{2}{c|}{} & $0.0002$ & $0.0003$ & $0.0005$ \\
\hline
$\sigma$ & \multicolumn{2}{c|}{$176805$} & $621.7$ & $944.8$ & $842.8$ \\
$N_{\ell}$ cut & \multicolumn{2}{c|}{$140132$} & $485.2$ & $726.0$ & $649.4$ \\
$2.2\;{\rm TeV}<E_{\ell}<22\;{\rm TeV}$ & \multicolumn{2}{c|}{$7786.7$} & $462.2$ & $688.2$ & $603.0$ \\
$2.8\;{\rm TeV}<m_{\ell\ell}<18\;{\rm TeV}$ & \multicolumn{2}{c|}{$3288.0$} & $366.8$ & $553.3$ & $477.1$  \\
$W_{\rm score}>0.85$ & $482.0$ & & & $362.5$ & $346.4$ \\
$W_{\rm score}>0.9$  & & $299.2$ & $246.2$ & &  \\
\end{tabular}
\end{center}
\caption{\label{tab.cutflow}Cross-sections~(ab) after cuts at $\sqrt{s}=30$ TeV.}
\end{table}

The interference has been neglected until now.
For the coefficients listed in Table~\ref{tab.cutflow}, the contributions of interference terms compared with those of NP squared terms are shown in Table~\ref{tab.interference1}.
One can see that the interference of $O_{T_i}$ operators is about $10\%$ of the squared term and should be taken into account.
However, the interference does not only come from the interference of $WW\to WW$ but also from other diagrams such as tri-boson diagrams, which are meant to be cut off.
Apart from that, the event selection strategy designed to cut off background should also suppress the interference contribution.
Therefore, taking $O_{T_2}$ as an example (whose interference is the greatest among all operators in study), we investigate the effect of event selection strategy on the interference.
After the event selection strategy, $\sigma _{\rm int}$ becomes $28.6\;{\rm ab}$, which is less than $10\%$ of $\sigma _{\rm NP}$ after event selection strategy is applied.

\begin{table}
\begin{center}
\begin{tabular}{c|c|c|c}
 & $f_{X}/\Lambda ^4$ & $\sigma _{\rm NP}$ & $\sigma _{\rm int}$ \\
 & $(10^{-3}\;{\rm TeV}^{-4})$ & & \\
\hline
$O_{S_0}$ & $100$ & $23559$ & $37.2$ \\
$O_{S_1}$ & $100$ & $56280$ & $-6.3$ \\
$O_{S_2}$ & $100$ & $56326$ & $-0.81$ \\
$O_{M_0}$ & $2$ & $964.0$ & $0.28$ \\
$O_{M_1}$ & $4$ & $714.2$ & $18.5$ \\
$O_{M_7}$ & $8$ & $955.5$ & $-23.3$ \\
$O_{T_0}$ & $0.2$ & $621.7$ & $47.7$ \\
$O_{T_1}$ & $0.3$ & $944.8$ & $76.5$ \\
$O_{T_2}$ & $0.5$ & $842.8$ & $93.5$ \\
\end{tabular}
\end{center}
\caption{\label{tab.interference1}Cross-sections~(ab) of interference terms and squared terms.}
\end{table}

%%%%%%%%%%%%%%%%%%%%%%%%%%%%%%%%%%%%%%%5
\subsection{\label{sec5.3}Signal significance and the expected constraints}
%%%%%%%%%%%%%%%%%%%%%%%%%%%%%%%%%%%%%%%

As introduced, we compare the cross-sections under a certain energy scale corresponding to the coefficients of dimension-8 operators.
As a result, due to the energy cut caused by unitarity bound depending on the coefficients, the cross-section of the SM appears to be functions of coefficients.
From another point of view, the cross-section of the SM does not actually depend on the coefficients of the operators, but on a certain energy scale that we have selected.

\begin{figure*}
\begin{center}
\includegraphics[width=0.32\textwidth]{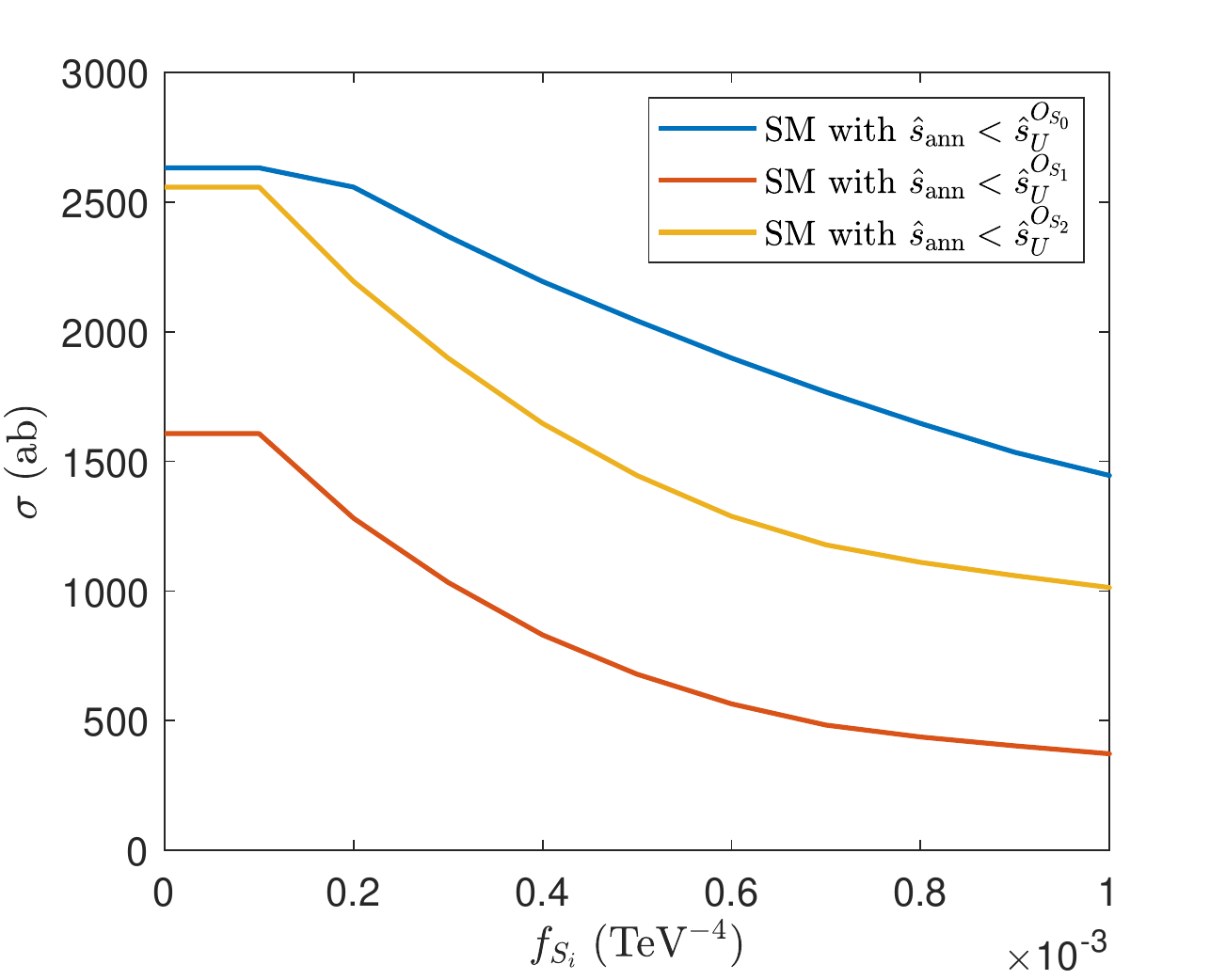}
\includegraphics[width=0.32\textwidth]{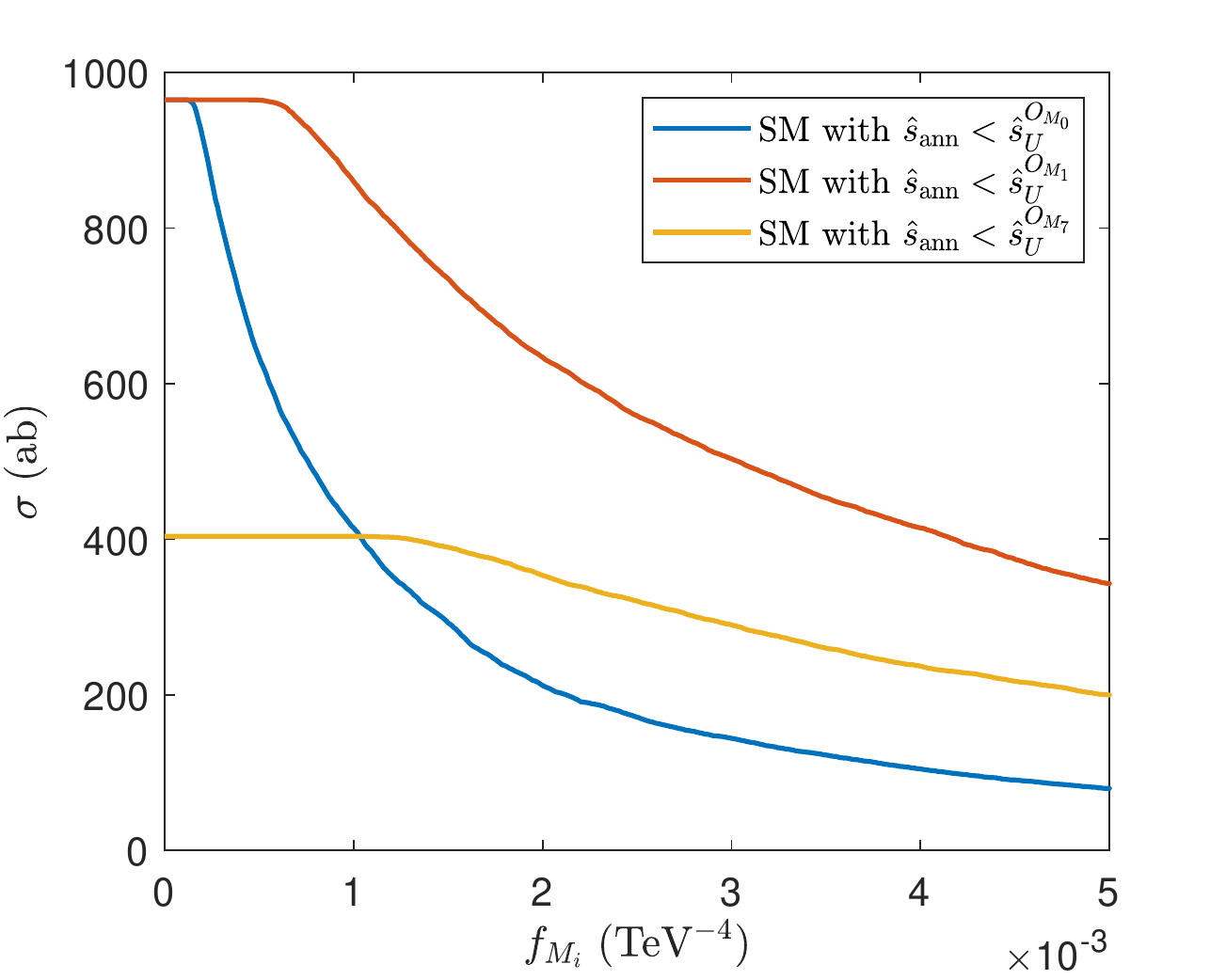}
\includegraphics[width=0.32\textwidth]{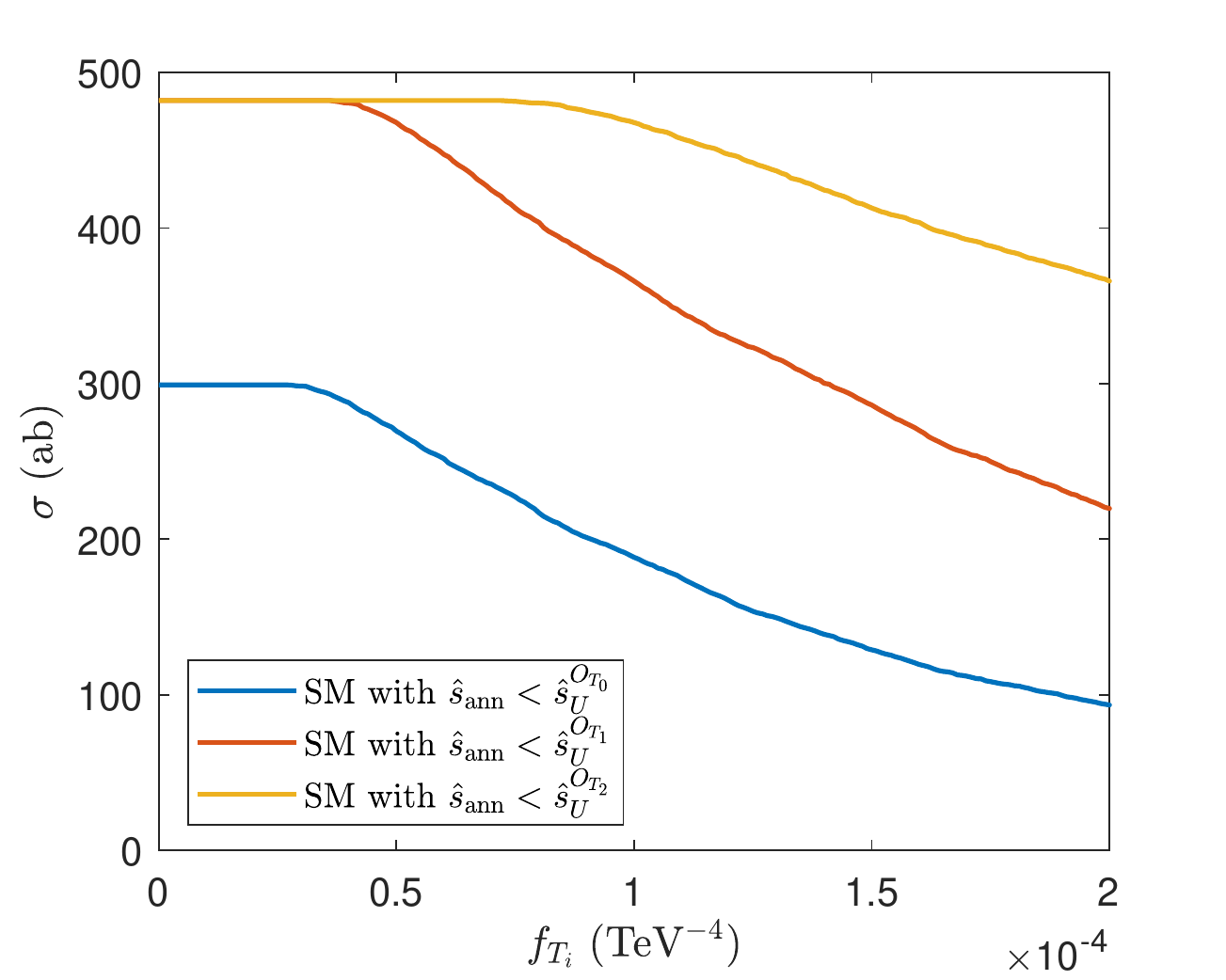}\\
\includegraphics[width=0.32\textwidth]{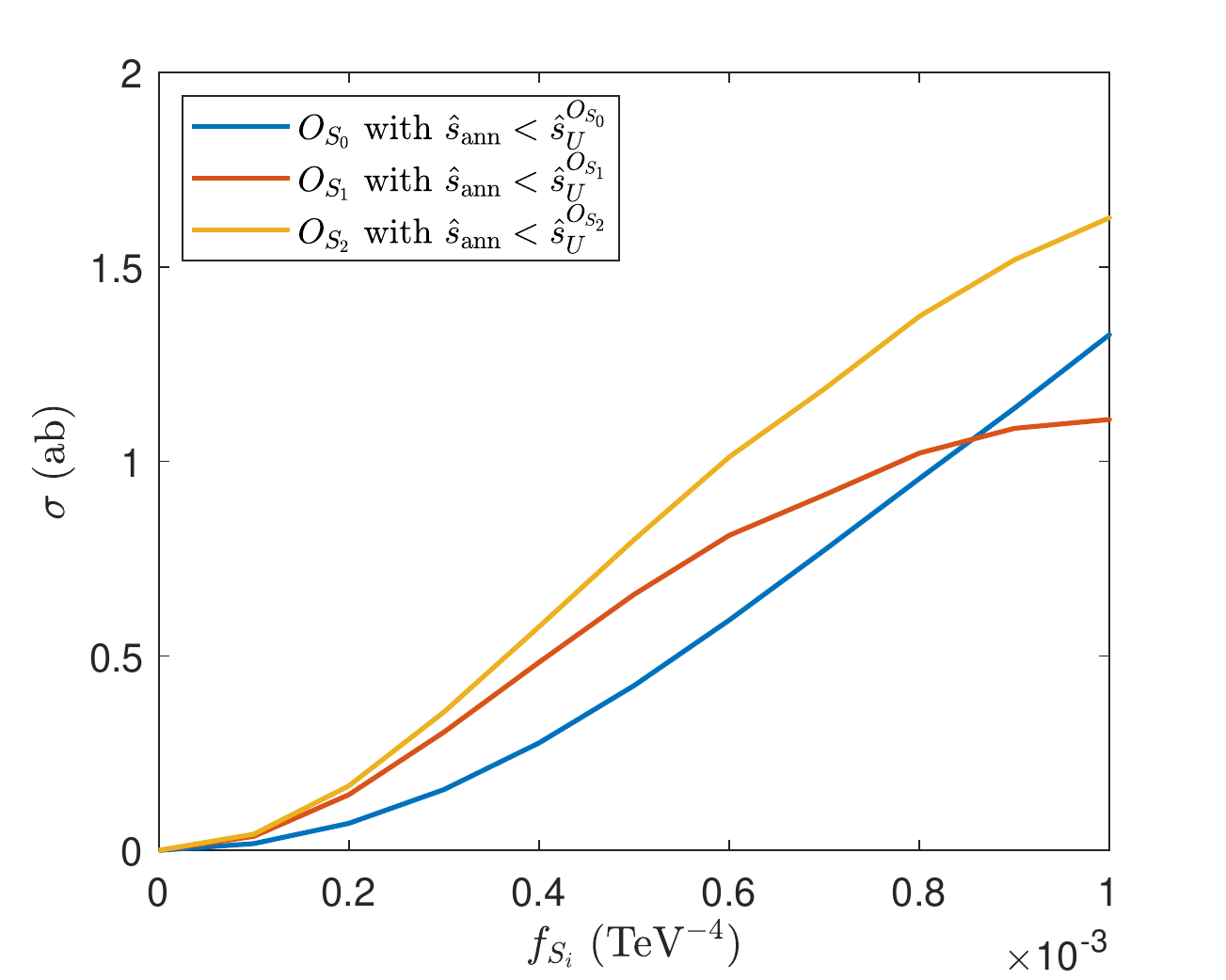}
\includegraphics[width=0.32\textwidth]{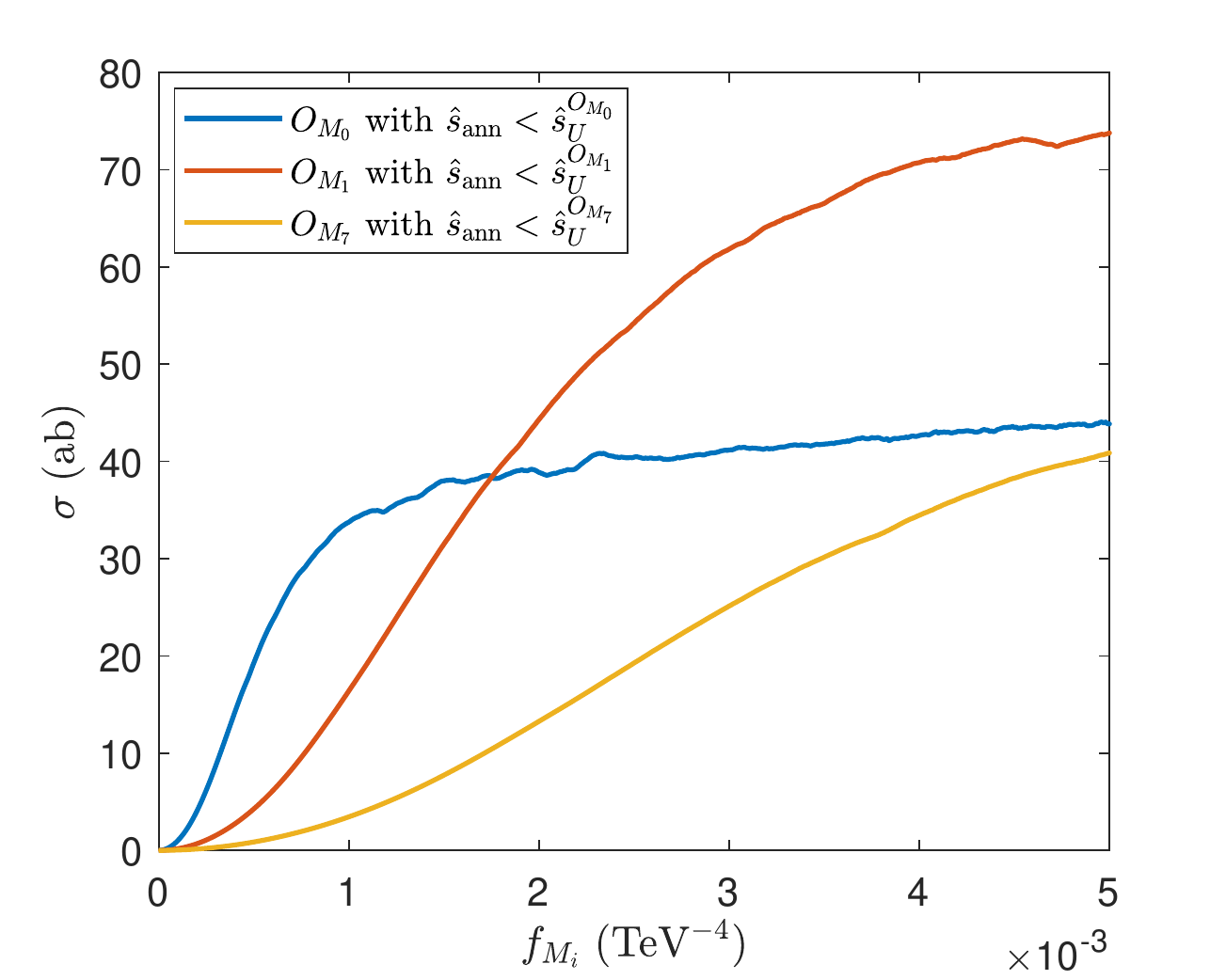}
\includegraphics[width=0.32\textwidth]{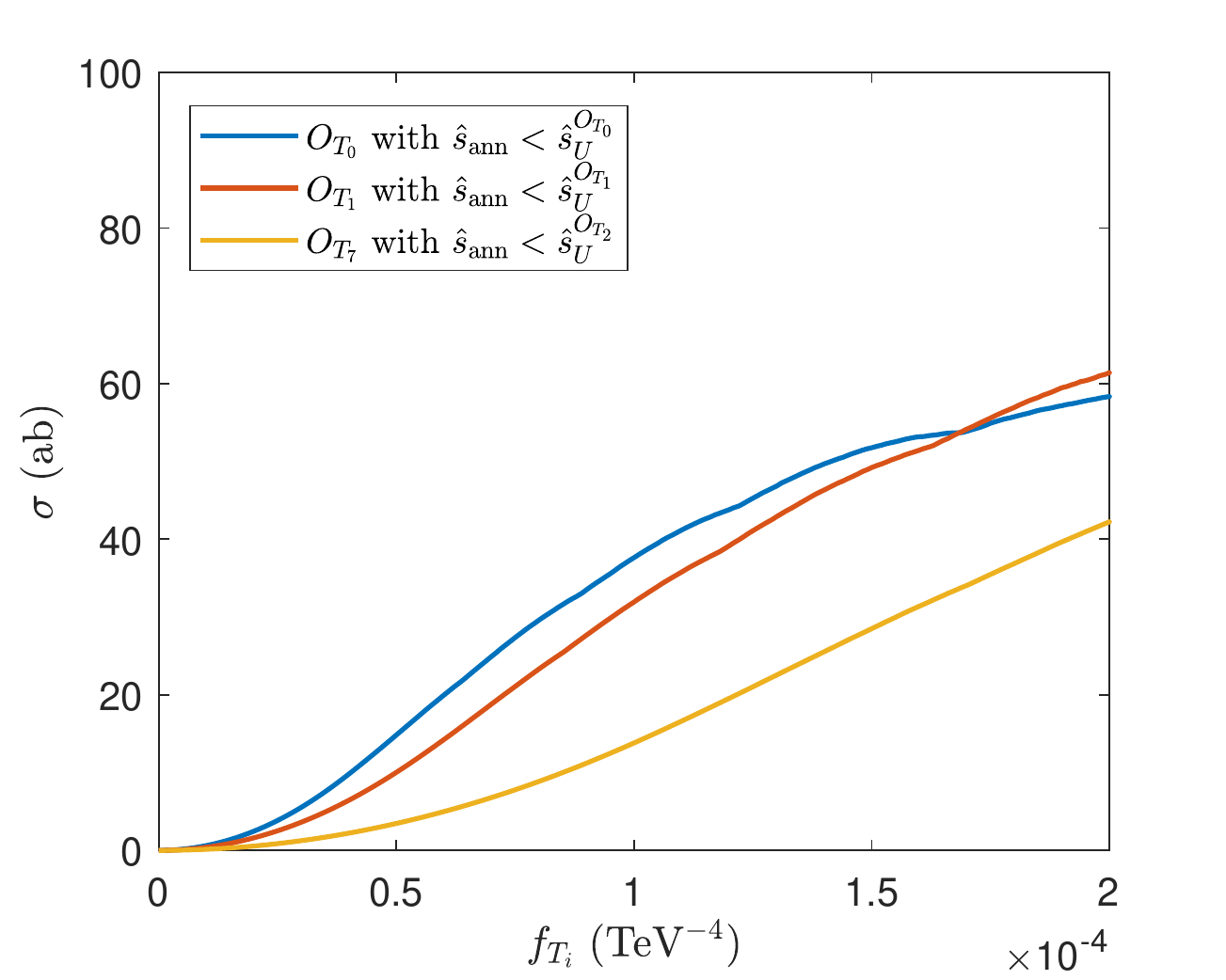}
\caption{\label{fig:crosssection}The cross-sections after unitarity bounds are applied.}
\end{center}
\end{figure*}

The cross-sections after applying unitarity bounds are shown in Fig.~\ref{fig:crosssection}.
The upper bounds of $\hat{s}$ are calculated using Eq.~(\ref{eq.5.4}) and denoted as $\hat{s}_U^{O_X}$ for each operator $O_X$.
It can be found that, in the ranges of coefficients considered in this paper, the cross-sections of the SM are typically $\sim \mathcal{O}(1)\;{\rm fb}$, and the cross-sections of NP are typically less than $0.1\;{\rm fb}$.

For luminosities $L=10\;{\rm ab}^{-1}$ and $L=90\;{\rm ab}^{-1}$~\cite{muoncollider5}, the expected constraints on the absolute values of coefficients~($|f_X|$) are obtained with the help of signal significance and assuming one operator at a time.
Since there are still errors between $\hat{s}_{\rm ann}$ and $\hat{s}_{\rm tr}$, moreover, the EFT probably stops being valid before the unitarity limit.
Therefore, the robustness of the results are studied by varying the cut-off on $\hat{s}_{\rm ann}^2$ by factors $1/2$ and $2$, analogous to what has been done with the QCD scales for the study of aQGCs at the LHC~\cite{zastudy}.
The results are listed in Tables~\ref{tab.const10} and \ref{tab.const90}.
The expected constraints with $2\hat{s}_U^2$ and $\hat{s}_U^2/2$ used as unitarity bounds are presented as systematic errors in Tables~\ref{tab.const10} and \ref{tab.const90}.

As introduced, the $O_{S_i}$ operators are significantly affected by the unitarity bounds, which can also be seen from Tables~\ref{tab.const10} and \ref{tab.const90}.
Take the expected constraints at $\mathcal{S}_{stat}=2$, for $O_{S_i}$ operators in general and $O_{M_i}$ operators at $L=10\;{\rm ab}^{-1}$, our results only show the orders of magnitude. For $O_{T_i}$ operators in general and $O_{M_i}$ operators at $L=90\;{\rm ab}^{-1}$, however, our results do not rely on the unitarity bounds and are thus more meaningful for experiments.
This is a representation of the ``EFT triangle'' problem~\cite{mo1,efttraingle2,efttraingle3,wwwwunitary}, which can be solved by high luminosity.
As a result, one can see that the results for $O_{M_i}$ and $O_{T_i}$ operators at $L=90\;{\rm ab}^{-1}$ are more reliable.
Generally, the $O_{T_i}$ operators are seldom affected by the unitarity bounds. This is because the luminosity of the transverse $W$ bosons from the beam is logarithmically enhanced, and all the dominant helicity amplitudes of $O_{T_i}$ contain transverse $W$ bosons in the initial state of $WW\to WW$.
Compared with Table~\ref{tab.1} which does not consider unitarity bounds, the coefficients can be narrowed down significantly even with unitarity bounds considered. The expected constraints can be $3$ to $4$ orders of magnitude stronger than those at the $13\;{\rm TeV}$ LHC for $O_{M_i}$ and $O_{T_i}$ operators.

\begin{table}
\begin{center}
\begin{tabular}{c|c|c|c}
$\mathcal{S}_{stat}$ & $f_{S_0}/\Lambda^4$ & $f_{S_1}/\Lambda^4$ & $f_{S_2}/\Lambda^4$ \\
\hline
$2$ & $35.3^{+39.5}_{-27.5}$ & $61.4^{+21.8}_{-15.6}$ & $42.4^{+10.8}_{-0.3}$ \\
$3$ & $72.0^{+20.7}_{-59.5}$ & $78.0^{+30.0}_{-5.2}$ & $60.0^{+14.6}_{-1.0}$ \\
$5$ & $111.7^{+48.0}_{-87.7}$ & $166.3^{+48.6}_{-52.9}$ & $95.8^{+12.5}_{-4.3}$ \\
\hline
$\mathcal{S}_{stat}$ & $f_{M_0}/\Lambda^4$ & $f_{M_1}/\Lambda^4$ & $f_{M_7}/\Lambda^4$ \\
\hline
$2$ & $0.45^{+0.59}_{-0.01}$ & $1.08^{+0.31}_{-0}$ & $1.91^{+0.34}_{-0}$ \\
$3$ & $0.59^{+1.28}_{-0.05}$ & $1.35^{+1.07}_{-0.03}$ & $2.37^{+1.31}_{-0}$ \\
$5$ & $0.99^{+4.15}_{-0.29}$ & $1.90^{+3.88}_{-0.19}$ & $3.23^{+8.92}_{-0.15}$ \\
\hline
$\mathcal{S}_{stat}$ & $f_{T_0}/\Lambda^4$ & $f_{T_1}/\Lambda^4$ & $f_{T_2}/\Lambda^4$ \\
\hline
$2$ & $0.043^{+0.006}_{-0}$ & $0.059^{+0.010}_{-0.001}$ & $0.101^{+0.012}_{-0}$ \\
$3$ & $0.053^{+0.014}_{-0}$ & $0.073^{+0.023}_{-0.001}$ & $0.124^{+0.035}_{-0}$ \\
$5$ & $0.072^{+0.044}_{-0.004}$ & $0.100^{+0.069}_{-0.005}$  & $0.166^{+0.109}_{-0.004}$ \\
\end{tabular}
\end{center}
\caption{\label{tab.const10}The expected constraints on the absolute values of coefficients~($10^{-3}\;{\rm TeV}^{-4}$) at $\sqrt{s}=30$ TeV and $L=10\;{\rm ab}^{-1}$.
The results by varying the unitarity bounds by factors $1/2$ and $2$ are presented as systematic errors.}
\end{table}

\begin{table}
\begin{center}
\begin{tabular}{c|c|c|c}
$\mathcal{S}_{stat}$ & $f_{S_0}/\Lambda^4$ & $f_{S_1}/\Lambda^4$ & $f_{S_2}/\Lambda^4$ \\
\hline
 $2$ & $5.3^{+21.1}_{-2.6}$ & $31.3^{+0.8}_{-29.5}$ & $5.9^{+19.4}_{-3.8}$ \\
 $3$ & $8.7^{+34.6}_{-5.0}$ & $40.8^{+1.4}_{-36.8}$ & $22.9^{+9.3}_{-19.6}$ \\
 $5$ & $18.2^{+41.7}_{-11.9}$ & $56.9^{+23.0}_{-17.9}$ & $35.0^{+13.8}_{-29.1}$ \\
\hline
$\mathcal{S}_{stat}$ & $f_{M_0}/\Lambda^4$ & $f_{M_1}/\Lambda^4$ & $f_{M_7}/\Lambda^4$ \\
\hline
 $2$ & $0.26^{+0.01}_{-0}$ & $0.62^{+0.01}_{-0}$ & $1.12^{+0.01}_{-0}$ \\
 $3$ & $0.31^{+0.07}_{-0}$ & $0.76^{+0.04}_{-0}$ & $1.36^{+0.02}_{-0}$ \\
 $5$ & $0.40^{+0.32}_{-0}$ & $0.98^{+0.19}_{-0}$ & $1.75^{+0.19}_{-0}$ \\
\hline
$\mathcal{S}_{stat}$ & $f_{T_0}/\Lambda^4$ & $f_{T_1}/\Lambda^4$ & $f_{T_2}/\Lambda^4$ \\
\hline
 $2$ & $0.025^{+0}_{-0}$ & $0.034^{+0}_{-0}$ & $0.058^{+0}_{-0}$ \\
 $3$ & $0.030^{+0.001}_{-0}$ & $0.042^{+0.001}_{-0}$ & $0.072^{+0}_{-0}$ \\
 $5$ & $0.039^{+0.003}_{-0}$ & $0.054^{+0.006}_{-0}$ & $0.092^{+0}_{-0}$ \\
\end{tabular}
\end{center}
\caption{\label{tab.const90}The expected constraints on the absolute values of coefficients~($10^{-3}\;{\rm TeV}^{-4}$) at $\sqrt{s}=30$ TeV and $L=90\;{\rm ab}^{-1}$.
The results by varying the unitarity bounds by factors $1/2$ and $2$ are presented as systematic errors.}
\end{table}

\section{\label{sec6}Summary}

As a gauge boson collider, the muon collider is suitable to study the VBS processes for the aQGCs.
Unlike the LHC, with cleaner final states, it is easier to extract aQGCs out of all relevant dimension-8 operators.
For example, exclusive $W^+W^-\to W^+W^-$ can be separated out from $VV\to W^+W^-$ processes.
In this paper, we study the process $\mu^+\mu^-\to \ell^+\ell^-\nu\nu\bar{\nu}\bar{\nu}$ containing a $W^+W^-\to W^+W^-$ subprocess induced by dimension-8 operators contributing to aQGCs.

The presence of four (anti-)neutrinos in the final states poses difficulties to the phenomenological study.
In this paper, this problem is tackled with a machine learning approach.
The ANNs can be used to pick out the contribution including a $W^+W^-\to W^+W^-$ subprocess.
Moreover, the ANNs can also be used to reconstruct $\hat{s}$ of the $W^+W^-\to W^+W^-$ subprocess.

With the help of the ANNs, the sensitivities of the process $\mu^+\mu^-\to \ell^+\ell^-\nu\nu\bar{\nu}\bar{\nu}$ to the dimension-8 operators contributing to aQGCs are studied at muon collider with $\sqrt{s}=30$ TeV.
The kinematic features are investigated, and the event selection strategies are proposed.
The unitarity bounds are also considered, which turn out to be necessary in the study of the $W^+W^-\to W^+W^-$ subprocess at a muon collider.
The expected constraints are studied with the help of signal significance, and are found to be as small as $4$ orders of magnitude stronger than those at the $13\;{\rm TeV}$ LHC even with unitarity bounds applied.
This shows the great advantage of the muon collider in studying the aQGCs.

\section*{ACKNOWLEDGMENT}

\noindent
This work was supported in part by the National Natural Science Foundation of China under Grants Nos. 11905093 and 12147214, the Natural Science Foundation of the Liaoning Scientific Committee No.~LJKZ0978 and the Outstanding Research Cultivation Program of Liaoning Normal University (No.21GDL004).
T.L. is supported by the National Natural Science Foundation of China (Grants No. 12035008, 11975129) and ``the Fundamental Research Funds for the Central Universities'', Nankai University (Grant No. 63196013).

\appendix

\section{\label{ap1}Effective vector boson approximation}

As a comparison, the contribution of NP involving a $W^+W^-\to W^+W^-$ subprocess is calculated by using effective vector boson approximation.
The $W^+W^-\to W^+W^-$ contribution to the process $\mu^+\mu^-\to \nu\nu\bar{\nu}\bar{\nu}\ell^+\ell^-$ can be given by EVA as~\cite{eva1,eva2,eva3}
\begin{equation}
\begin{split}
&\sigma_{\rm VBS} (\mu^+\mu^-\to \bar{\nu}\nu W^+W^-)=\sum _{\lambda _1\lambda _2\lambda _3\lambda _4}\int d\xi _1\int d\xi_2 f_{W_{\lambda _1}^-/\mu^-}(\xi _1)f_{W_{\lambda _2}^+/\mu^+}(\xi _2)\sigma _{W_{\lambda _1}^+W_{\lambda _2}^-\to W^+_{\lambda _3}W^-_{\lambda _4}}(\hat{s}),\\
&f_{W_{+1}^-/\mu _L^-}(\xi)=f_{W_{-1}^+/\mu_L^+}(\xi)=\frac{e^2}{8\pi^2s_W^2}\frac{(1-\xi)^2}{2\xi}\log \frac{\mu_f^2}{M_W},\\
&f_{W_{-1}^-/\mu _L^-}(\xi)=f_{W_{+1}^+/\mu_L^+}(\xi)=\frac{e^2}{8\pi^2s_W^2}\frac{1}{2\xi}\log \frac{\mu_f^2}{M_W},\\
&f_{W_{0}^-/\mu _L^-}(\xi)=f_{W_{0}^+/\mu_L^+}(\xi)=\frac{e^2}{8\pi^2s_W^2}\frac{1-\xi}{\xi},\\
&f_{W_{\lambda}^{\pm}/\mu _R^{\pm}}=0,\;\;\;\;f_{W_{\lambda}^{\pm}/\mu _R^{\pm}}=\frac{f_{W_{\lambda}^{\pm}/\mu _L^{\pm}}+f_{W_{\lambda}^{\pm}/\mu _R^{\pm}}}{2},\\
\end{split}
\label{eq.ap.1}
\end{equation}
where $\sqrt{\hat{s}}=\sqrt{\xi _1\xi _2 s}$ and $\mu _f$ is the factorization scale which can be set to be $\sqrt{\hat{s}}/4$~\cite{eva3}. At the leading order of $s$, one has
\begin{equation}
\begin{split}
&\sigma_{\rm VBS} (\mu^+\mu^-\to \bar{\nu}\nu W^+W^-)=\frac{e^4 s^3}{16986931200000 \pi ^5 s_W^4\Lambda^8} \\
&\times \left\{40 \log \left(\frac{s}{16 M_W^2}\right) \left[17385600 f_{M_0}^2+2400 f_{M_0} (2759 f_{M_7}-3622 f_{M_1})-1154484 f_{M_1}^2\right.\right.\\
&\left.\left.+585684 f_{M_1} f_{M_7}-134871  f_{M_7}^2+128 \left[7500 \left(6  f_{S_0}^2+11 f_{S_0} (f_{S_1}+f_{S_2})+14 (f_{S_1}+f_{S_2})^2\right)\right.\right.\right.\\
&\left.\left.\left.+3744668 f_{T_0}^2+4574508 f_{T_0} f_{T_1}+2688508 f_{T_0} f_{T_2}+2078155 f_{T_1}^2+2052462 f_{T_1} f_{T_2}+665294 f_{T_2}^2\right]\right.\right.\\
&\left.\left.+15 \log \left(\frac{s}{16 M_W^2}\right) \left(9600 f_{M_0}^2+2400 f_{M_0} (f_{M_7}-2 f_{M_1})+1004 f_{M_1}^2-1004 f_{M_1} f_{M_7}+301 f_{M_7}^2\right.\right.\right.\\
&\left.\left.\left.+128 \left(4652 f_{T_0}^2+5532 f_{T_0} f_{T_1}+3532 f_{T_0} f_{T_2}+3055 f_{T_1}^2+3198 f_{T_1} f_{T_2}+1046 f_{T_2}^2\right)\right)\right.\right.\\
&\left.\left.-163200 f_{M_0}^2+600 f_{M_0} (136 f_{M_1}-167 f_{M_7})+47612 f_{M_1}^2-32762 f_{M_1} f_{M_7}+9403 f_{M_7}^2\right.\right.\\
&\left.\left.-128 \left(139804 f_{T_0}^2+169284 f_{T_0} f_{T_1}+102284 f_{T_0} f_{T_2}+82295 f_{T_1}^2+83076 f_{T_1} f_{T_2}+27022 f_{T_2}^2\right)\right]\right\}.
\end{split}
\label{eq.ap.2}
\end{equation}
With ${\rm Br}(W\to e\nu_e)\approx {\rm Br}(W\to \mu\nu_{\mu})\approx 10.8\%$~\cite{pdg}, the predictions of $\sigma _{\rm 4W}$ by Eq.~(\ref{eq.ap.2})~(denoted as $\sigma ^{\rm EVA}_{\rm 4W}$) can be calculated.

The differential cross-section is also calculated to estimate the effect of unitarity bounds, which is
\begin{equation}
\begin{split}
&\frac{d\sigma _{\rm 4W}(\mu^+\mu^-\to \bar{\nu}\nu W^+W^-)}{d\hat{s}}=-\frac{e^4 \hat{s}^2 }{31457280 \pi ^5 \Lambda^8s^2 s_W^4}\left\{8 s (A_0+A_2) \left(2 (s-\hat{s})+(s+\hat{s}) \log \left(\frac{\hat{s}}{s}\right)\right)\right.\\
&\left.+\log ^2\left(\frac{\hat{s}}{16 M_W^2}\right) \left[(s-\hat{s}) (\hat{s} (-(-4 A_1+A_2+A_3)))+\left(A_1 \left(4 s \hat{s}+\hat{s}^2\right)\right) \log \left(\frac{\hat{s}}{s}\right)\right.\right.\\
&\left.\left.+(A_2+A_3) \left(2 s^2 \log \left(\frac{\hat{s}}{s}\right)+3 s (s-\hat{s})\right)\right]\right.\\
&\left.+A_4 \log \left(\frac{\hat{s}}{16 M_W^2}\right) \left((s-\hat{s}) (7 s+\hat{s})+4 s (s+\hat{s}) \log \left(\frac{\hat{s}}{s}\right)\right)\right\},\\
\end{split}
\label{eq.ap.3}
\end{equation}
where
\begin{equation}
\begin{split}
&A_0=32 \left(6 f_{S_0}^2+11 f_{S_0} (f_{S_1}+f_{S_2})+14 (f_{S_1}+f_{S_2})^2\right),\\
&A_1=(f_{M_7}-2 f_{M_1})^2+384 \left(4 f_{T_0}^2+4 f_{T_0} (f_{T_1}+f_{T_2})+5 f_{T_1}^2+6 f_{T_1} f_{T_2}+2 f_{T_2}^2\right),\\
&A_2=480 f_{M_0}^2+120 f_{M_0} (f_{M_7}-2 f_{M_1})+34 f_{M_1}^2-34 f_{M_1} f_{M_7}+11 f_{M_7}^2,\\
&A_3=128 \left(184 f_{T_0}^2+228 f_{T_0} f_{T_1}+128 f_{T_0} f_{T_2}+92 f_{T_1}^2+87 f_{T_1} f_{T_2}+28 f_{T_2}^2\right),\\
&A_4=2 \left(96 f_{M_0}^2-12 f_{M_0} (4 f_{M_1}+f_{M_7})+46 f_{M_1}^2-37 f_{M_1} f_{M_7}+11 f_{M_7}^2\right).\\
\end{split}
\label{eq.ap.4}
\end{equation}
The cross-section with unitarity bounds is then estimated as
\begin{equation}
\begin{split}
&\sigma _{\rm 4W,U}=\int _0^{\min\{\hat{s}_{\rm max},s\}}d\hat{s}\frac{d\sigma _{\rm 4W}}{d\hat{s}},
\end{split}
\label{eq.ap.5}
\end{equation}
where $\hat{s}_{\rm max}$ is the maximally allowed $\hat{s}$ in the sense of unitarity.
Assuming one operator at a time, $\hat{s}_{\rm max}$ is given in Eq.~(\ref{eq.5.4}).
As an example to illustrate the effect of unitarity bounds, consider only the existence of $O_{S_0}$,
\begin{equation}
\begin{split}
&\sigma _{\rm 4W}(\mu^+\mu^-\to \bar{\nu}\nu W^+W^-)=\frac{e^4 f_{S_0}^2 s^3}{2949120 \pi ^5 \Lambda^8s_W^4},\\
&\sigma _{\rm 4W, U}(\mu^+\mu^-\to \bar{\nu}\nu W^+W^-)=\frac{e^4 f_{S_0}^2 s^3(1-B_{S_0})^3}{2949120 \pi ^5 s_W^4}\\
&\times \left[1+9 B_{S_0} \left(4 \log \left(1-B_{S_0}\right)-9\right)-84 \log \left(1-B_{S_0}\right)\right],\\
&B_{S_0}=\left(1-\frac{4 \sqrt{3 \pi }}{s\sqrt{|f_{S_0}|}}\right) \theta \left(s-\frac{4 \sqrt{3 \pi }}{\sqrt{|f_{S_0}|}}\right),\\
\end{split}
\label{eq.ap.6}
\end{equation}
where $\sigma _{\rm 4W,U}(\mu^+\mu^-\to \ell ^+\ell^-\bar{\nu}\bar{\nu}\nu\nu)=\sigma _{\rm 4W,U}(\mu^+\mu^-\to \bar{\nu}\nu W^+W^-)\times ({\rm Br}(W\to \ell \nu))^2$, $\sigma _{\rm 4W,U}(\mu^+\mu^-\to \ell ^+\ell^-\bar{\nu}\bar{\nu}\nu\nu)$ denotes the cross-section of $W^+W^-\to W^+W^-$ induced $\mu^+\mu^-\to \ell^+\ell^-\nu\nu\bar{\nu}\bar{\nu}$ with unitarity bound applied, $\theta(x)$ is the Heaviside unit step function.
Similarly,
\begin{equation}
\begin{split}
&\sigma^{O_{S_{1,2}}} _{\rm 4W, U}(\mu^+\mu^-\to \bar{\nu}\nu W^+W^-)=\frac{7 e^4 f_{S_{1,2}}^2s^3(1-B_{S_{1,2}})^3}{8847360 \pi ^5 s_W^4} \\
&\times \left[1+9 B_{S_{1,2}} \left(4 \log \left(1-B_{S_{1,2}}\right)-9\right)-84 \log \left(1-B_{S_{1,2}}\right)\right],\\
\end{split}
\label{eq.ap.7}
\end{equation}
\begin{equation}
\begin{split}
&\sigma^{O_{M_0}} _{\rm 4W, U}(\mu^+\mu^-\to \bar{\nu}\nu W^+W^-)=\frac{(B_{M_0}-1)^3 e^4 s^3 f_{M_0}^2 }{1769472000 \pi ^5 s_W^4} \left\{122500 \log (1-B_{M_0})\right.\\
&\left.-\frac{864 \sqrt{2 \pi } B_{M_0}}{s\sqrt{|f_{M_0}|}} \left[2-5 \log \left(\frac{(1-B_{M_0}) s}{16 M_W^2}\right)\right]^2-51300 B_{M_0} \log (1-B_{M_0})\right.\\
&\left.+600 [36 B_{M_0}+30 \log (1-B_{M_0})-1] \log ^2\left(\frac{(1-B_{M_0}) s}{16 M_W^2}\right)+118611 B_{M_0}-1811\right.\\
&\left.+40 [243 B_{M_0}+30 (11-9 B_{M_0}) \log (1-B_{M_0})+17] \log \left(\frac{(1-B_{M_0}) s}{16 M_W^2}\right)\right\},
\end{split}
\label{eq.ap.8}
\end{equation}
\begin{equation}
\begin{split}
&\sigma^{O_{M_1}} _{\rm 4W, U}(\mu^+\mu^-\to \bar{\nu}\nu W^+W^-)=\frac{(B_{M_1}-1)^3 e^4 s^3 f_{M_1}^2 }{4246732800000 \pi ^5 s_W^4 \left(\frac{8 \sqrt{2 \pi }}{s\sqrt{|f_{M_1}|}}-1\right)}\left\{-9 (1607931 B_{M_1}+32069)\right.\\
&\left.+600 \log ^2\left(\frac{(1-B_{M_1}) s}{16 M_W^2}\right) \left((251-6201 B_{M_1})+\frac{8 \sqrt{2 \pi } (6975 B_{M_1}-251)}{s\sqrt{|f_{M_1}|}} -\frac{99072 \pi  B_{M_1}}{s^2|f_{M_1}|}\right.\right.\\
&\left.\left.-60 \log (1-B_{M_1}) \left((103-18 B_{M_1})+\frac{8 \sqrt{2 \pi } (15 B_{M_1}-103)}{s\sqrt{|B_{M_1}|}}+\frac{384 \pi  B_{M_1}}{s^2|f_{M_1}|}\right)\right)\right.\\
&\left.+40 \log \left(\frac{(1-B_{M_1}) s}{16 M_W^2}\right) \left((11903-555903 B_{M_1})+\frac{8 \sqrt{2 \pi } (489375 b-11903)}{s\sqrt{|B_{M_1}|}}+\frac{8515584 \pi  B_{M_1}}{s^2|f_{M_1}|}\right.\right.\\
&\left.\left.+30 \log (1-B_{M_1}) \left((10089 B_{M_1}-22189) +\frac{8 \sqrt{2 \pi } (22189-10125 B_{M_1})}{s\sqrt{|B_{M_1}|}} +\frac{4608 \pi  B_{M_1}b}{s^2|f_{M_1}|}\right)\right)\right.\\
&\left.-20 \log (1-B_{M_1}) \left((677557-307557 B_{M_1}) +\frac{8 \sqrt{2 \pi } (307125 B_{M_1}-677557)}{s\sqrt{|B_{M_1}|}}+\frac{55296 \pi  B_{M_1}}{s^2|f_{M_1}|}\right)\right.\\
&\left.+\frac{72 \sqrt{2 \pi } (1666875 B_{M_1}+32069)}{s\sqrt{|B_{M_1}|}}-\frac{67903488 \pi  B_{M_1}}{s^2|f_{M_1}|}\right\},
\end{split}
\label{eq.ap.9}
\end{equation}
\begin{equation}
\begin{split}
&\sigma^{O_{M_7}} _{\rm 4W, U}(\mu^+\mu^-\to \bar{\nu}\nu W^+W^-)=\frac{(B_{M_7}-1)^3 e^4 s^3 f_{M_7}^2}{16986931200000 \pi ^5 s_W^4} \left\{33 (635913 B_{M_7}+4087)\right.\\
&\left.-600 \log ^2\left(\frac{(1-B_{M_7}) s}{16 M_W^2}\right) \left[9 B_{M_7} \left(\frac{2176 \sqrt{\pi }}{s\sqrt{|f_{M_7}|}}-889\right)\right.\right.\\
&\left.\left.+120 \log (1-B_{M_7}) \left((9 B_{M_7}-64)+\frac{24 \sqrt{\pi } B_{M_7}}{s\sqrt{|f_{M_7}|}}\right)+301 \right]\right.\\
&\left.+40 \log \left(\frac{(1-B_{M_7}) s}{16 M_W^2}\right) \left[189 B_{M_7}\left(2727+\frac{5632 \sqrt{\pi }}{s\sqrt{|f_{M_7}|}}\right)\right.\right.\\
&\left.\left.+30 \log (1-B_{M_7}) \left(-9639 B_{M_7} +\frac{576 \sqrt{\pi } B_{M_7}}{s\sqrt{|f_{M_7}|}}+20639 \right)-9403 \right]\right.\\
&\left.-20 \log (1-B_{M_7}) \left((449307 b-1021307)+\frac{6912 \sqrt{\pi } B_{M_7}}{s\sqrt{|f_{M_7}|}}\right)-\frac{8487936 \sqrt{\pi } B_{M_7}}{s\sqrt{|f_{M_7}|}}\right\},
\end{split}
\label{eq.ap.10}
\end{equation}
\begin{equation}
\begin{split}
&\sigma^{O_{T_0}} _{\rm 4W, U}(\mu^+\mu^-\to \bar{\nu}\nu W^+W^-)=\frac{(B_{T_0}-1)^3 e^4 s^3 f_{T_0}^2}{33177600000 \pi ^5 s_W^4 \left(\frac{\sqrt{6 \pi }}{s\sqrt{|f_{T_0}|}}-1\right)} \left\{-\frac{1669248 \pi  B_{T_0}}{s^2|f_{T_0}|}\right.\\
&\left.+600 \log ^2\left(\frac{(1-B_{T_0}) s}{16 M_W^2}\right) \left[(1163-33363 B_{T_0}) +\frac{\sqrt{6 \pi } (39375 B_{T_0}-1163)}{s\sqrt{|f_{T_0}|}}\right.\right.\\
&\left.\left.-60 \log (1-B_{T_0}) \left(2 (257-27 B_{T_0}) +\frac{\sqrt{6 \pi } (45 B_{T_0}-514)}{s\sqrt{|f_{T_0}|}}+\frac{54 \pi  B_{T_0}}{s^2|f_{T_0}|}\right)-\frac{36072 \pi  B_{T_0}}{s^2|f_{T_0}|}\right]\right.\\
&\left.+40 \log \left(\frac{(1-B_{T_0}) s}{16 M_W^2}\right) \left[(264951 B_{T_0}-34951) +\frac{\sqrt{6 \pi } (34951-300375 B_{T_0})}{s\sqrt{|f_{T_0}|}}\right.\right.\\
&\left.\left.+30 \log (1-B_{T_0}) \left((9983-783 B_{T_0})+\frac{\sqrt{6 \pi } (675 B_{T_0}-9983)}{s\sqrt{|f_{T_0}|}}+\frac{648 \pi  B_{T_0}}{s^2|f_{T_0}|}\right)+\frac{212544 \pi  B_{T_0}}{s^2|f_{T_0}|}\right]\right.\\
&\left.+(936167-2776167 B_{T_0}) +\frac{\sqrt{6 \pi } (3054375 B_{T_0}-936167)}{s\sqrt{|f_{T_0}|}}\right.\\
&\left.-20 \log (1-B_{T_0}) \left(81 B_{T_0} \left(\frac{125 \sqrt{6 \pi }}{s\sqrt{|f_{T_0}|}}-141+\frac{96 \pi}{s^2|f_{T_0}|} \right)+195421 \left(1-\frac{\sqrt{6 \pi }}{s\sqrt{|f_{T_0}|}}\right)\right)\right\},
\end{split}
\label{eq.ap.11}
\end{equation}
\begin{equation}
\begin{split}
&\sigma^{O_{T_1}} _{\rm 4W, U}(\mu^+\mu^-\to \bar{\nu}\nu W^+W^-)=\frac{(B_{T_1}-1)^3 e^4 s^3 f_{T_1}^3 }{26542080000 \pi ^5 s_W^4 \left(\frac{2 \sqrt{2 \pi }}{s\sqrt{|f_{T_1}|}}-1\right)}\left\{415631  \left(1-\frac{2 \sqrt{2 \pi }}{s\sqrt{|f_{T_1}|}}\right)\right.\\
&\left.+600 \log ^2\left(\frac{(1-B_{T_1}) s}{16 M_W^2}\right) \left[(611-13491 B_{T_1}) + \frac{2 \sqrt{2 \pi } (14535 B_{T_1}-611)}{s\sqrt{|f_{T_1}|}} \right.\right.\\
&\left.\left.-120 \log (1-B_{T_1}) \left((119-27 B_{T_1}) +\frac{\sqrt{2 \pi } (45 B_{T_1}-238)}{s\sqrt{|f_{T_1}|}} +\frac{36 \pi  B_{T_1}}{s^2|f_{T_1}|}\right)-\frac{8352 \pi  B_{T_1}}{s^2|f_{T_1}|}\right]\right.\\
&\left.+40 \log \left(\frac{(1-B_{T_1}) s}{16 M_W^2}\right) \left[30 \log (1-B_{T_1}) \left((4463-783 B_{T_1}) + \frac{2 \sqrt{2 \pi } (675 B_{T_1}-4463)}{s\sqrt{|f_{T_1}|}} + \frac{864 \pi  B_{T_1}}{s^2|f_{T_1}|}\right)\right.\right.\\
&\left.\left.+351 B_{T_1} \left( 309 -\frac{650 \sqrt{2 \pi }}{s\sqrt{|f_{T_1}|}}+\frac{128 \pi}{s^2|f_{T_1}|} \right)-16459 \left(1-\frac{2 \sqrt{2 \pi }}{s\sqrt{|f_{T_1}|}}\right)\right]\right.\\
&\left.-27 B_{T_1} \left(42653-\frac{88250 \sqrt{2 \pi }}{s\sqrt{|f_{T_1}|}}+\frac{11776 \pi}{s^2|f_{T_1}|} \right)\right.\\
&\left.-20 \log (1-B_{T_1}) \left[81 B_{T_1} \left(\frac{250 \sqrt{2 \pi }}{s\sqrt{|f_{T_1}|}}-141 +\frac{128 \pi}{s^2|f_{T_1}|} \right)+85021\left(1-\frac{2 \sqrt{2 \pi }}{s\sqrt{|f_{T_1}|}}\right)\right]\right\},
\end{split}
\label{eq.ap.12}
\end{equation}
\begin{equation}
\begin{split}
&\sigma^{O_{T_2}} _{\rm 4W, U}(\mu^+\mu^-\to \bar{\nu}\nu W^+W^-)=\frac{(B_{T_2}-1)^3 e^4 s^3 f_{T_2}^2 }{66355200000 \pi ^5 s_W^4}
\left\{40 \log \left(\frac{(1-B_{T_2}) s}{16 M_W^2}\right) \left[-83511 B_{T_2} \right.\right.\\
&\left.\left.+30 \log (1-B_{T_2}) \left((783 B_{T_2}-3583) + \frac{432 \sqrt{\pi } B_{T_2}}{s\sqrt{|f_{T_2}|}}\right)+\frac{3456 \sqrt{\pi } B_{T_2}}{s\sqrt{|f_{T_2}|}}+13511 \right]\right.\\
&\left.-600 \log ^2\left(\frac{(1-B_{T_2}) s}{16 M_W^2}\right) \left[9 B_{T_2} \left(\frac{112 \sqrt{\pi }}{s\sqrt{|f_{T_2}|}}-1147\right)\right.\right.\\
&\left.\left.+120 \log (1-B_{T_2}) \left((27 B_{T_2}-97) + \frac{18 \sqrt{\pi } B_{T_2}}{s\sqrt{|f_{T_2}|}}\right)+523\right]\right.\\
&\left.+(27 B_{T_2} (64 B_{T_2}+32997)-332647) -20 \log (1-B_{T_2}) \left(81 B_{T_2} \left(141 +\frac{64 \sqrt{\pi }}{s\sqrt{|f_{T_2}|}}\right)-67421 \right)\right\},
\end{split}
\label{eq.ap.13}
\end{equation}
where
\begin{equation}
\begin{split}
&B_{S_{1,2}}=\left(1-\frac{2 \sqrt{6 \pi }}{s\sqrt{|f_{S_{1,2}}|}}\right) \theta \left(s-\frac{2 \sqrt{6 \pi }}{\sqrt{|f_{S_{1,2}}|}}\right),\;\;
 B_{M_0}=\left(1-\frac{4 \sqrt{2 \pi }}{s\sqrt{|f_{M_0}|}}\right) \theta \left(s-\frac{4 \sqrt{2 \pi }}{\sqrt{|f_{M_0}|}}\right),\\
&B_{M_1}=\left(1-\frac{8 \sqrt{2 \pi }}{\sqrt{|f_{M_1}|} s}\right) \theta \left(s-\frac{8 \sqrt{2 \pi }}{\sqrt{|f_{M_1}|}}\right),\;\;
 B_{M_7}=\left(1-\frac{16 \sqrt{2 \pi }}{\sqrt{|f_{M_7}|} s}\right) \theta \left(s-\frac{16 \sqrt{2 \pi }}{\sqrt{|f_{M_7}|}}\right),\\
&B_{T_0}=\left(1-\frac{\sqrt{6 \pi }}{s\sqrt{|f_{T_0}|}}\right) \theta \left(s-\frac{\sqrt{6 \pi }}{\sqrt{|f_{T_0}|}}\right),\;\;
 B_{T_1}=\left(1-\frac{2 \sqrt{2 \pi }}{s\sqrt{|f_{T_1}|}}\right) \theta \left(s-\frac{2 \sqrt{2 \pi }}{\sqrt{|f_{T_1}|}}\right),\\
&B_{T_2}=\left(1-\frac{4 \sqrt{\pi }}{s\sqrt{|f_{T_2}|}}\right) \theta \left(s-\frac{4 \sqrt{\pi }}{\sqrt{|f_{T_2}|}}\right).\\
\end{split}
\label{eq.ap.14}
\end{equation}

\section{\label{ap2}Helicity amplitudes relevant with the unitairty bounds}

With a large $\hat{s}$, we only need to focus on helicity amplitudes growing fastest with $\hat{s}$.
Denoting $\mathcal{M}(W_{\lambda _1}^+W_{\lambda _2}^-\to W_{\lambda _3}^+W_{\lambda _4}^-)=\mathcal{M}_{\lambda _1\lambda _2\lambda _3\lambda _4}+\mathcal{O}(\hat{s})$, the relevant amplitudes are
\begin{equation}
\begin{split}
\mathcal{M}_{++++}&=\frac{2f_{T_0}+f_{T_1}+f_{T_2}}{8\pi\Lambda^4}D_{0,0}^0,\\
\mathcal{M}_{++--}&=\frac{4f_{T_0}+3f_{T_1}+f_{T_2}}{12\pi\Lambda^4}D_{0,0}^0-\frac{4f_{T_0}-2f_{T_1}+f_{T_2}}{96\pi\Lambda^4}D_{0,0}^1+\frac{4f_{T_0}+6f_{T_1}+f_{T_2}}{480\pi\Lambda^4}D_{0,0}^2,\\
\mathcal{M}_{++00}&=-\frac{-8f_{M_0}-2f_{M_1}+f_{M_7}}{128\pi\Lambda^4}D_{0,0}^0-\frac{f_{M_7}}{384 \pi\Lambda^4}D_{0,0}^1,\\
\mathcal{M}_{+-+-}&=\frac{2f_{T_1}+f_{T_2}}{40\pi\Lambda^4}D_{2,2}^2,\;\;\;\;
\mathcal{M}_{+--+}=\frac{2f_{T_0}+f_{T_1}+f_{T_2}}{40\pi\Lambda^4}D_{2,-2}^2,\\
\mathcal{M}_{+-00}&=-\frac{2 f_{M_1}-f_{M_7}}{320 \sqrt{6} \pi\Lambda^4}D_{2,0}^2,\;\;\;\;
\mathcal{M}_{+0+0}=\frac{2 f_{M_1}-f_{M_7}}{192 \pi\Lambda^4 }D_{1,1}^1,\\
\mathcal{M}_{+0-0}&=\frac{-4 f_{M_0}+f_{M_1}+\frac{f_{M_7}}{3}}{256 \pi \Lambda^4}D_{1,-1}^1+\frac{4 f_{M_0}-f_{M_1}+f_{M_7}}{1280 \pi\Lambda^4 }D_{1,-1}^2,\\
\mathcal{M}_{0000}&=\frac{f_{S_0}+2 (f_{S_1}+f_{S_2})}{24 \pi \Lambda^4}D_{0,0}^0+\frac{2 f_{S_0}-f_{S_1}-f_{S_2}}{96 \pi\Lambda^4}D_{0,0}^1+\frac{2 f_{S_0}+f_{S_1}+f_{S_2}}{480 \pi\Lambda^4}D_{0,0}^2,\\
\end{split}
\label{eq.ap.15}
\end{equation}
where
\begin{equation}
\begin{split}
&D_{m_1,m_2}^J=\hat{s}^2 16\pi(J+\frac{1}{2})e^{i(m_1-m_2)\phi}d_{m_1,m_2}^J(\theta,\phi).\\
\end{split}
\label{eq.ap.16}
\end{equation}
The helicity amplitudes producing duplicated $T_J$ are not shown.

\bibliography{wwww}
\bibliographystyle{JHEP}

\end{document}